\documentclass[final,5p,times,twocolumn]{elsarticle}

\usepackage{amssymb}
\usepackage{amsthm}

\usepackage{amsmath,amsfonts}
\usepackage{siunitx}
\usepackage{caption}
\usepackage{subcaption}
\usepackage{mathrsfs}
\usepackage{color}
\graphicspath{{./figures/}}
\newcommand{\cp}{C_\mathrm{p}}

\newcommand{\fh}{\Delta h^{\circ}_{\mathrm{f},\alpha}}

\newcommand{\filt}[1]{\widetilde{#1}}
\newcommand{\rfilt}[1]{\overline{#1}}
\newcommand{\rfluc}[1]{{#1}'}
\newcommand{\fluc}[1]{{#1}''}

\newcommand{\di}{\,\mathrm{d}}
\newcommand{\lp}{\left(}
\newcommand{\rp}{\right)}

\newcommand{\nl}{ \notag \\ && &}

\newcommand{\pdt}[2][]{\partial_{#1}\lp{#2}\rp}

\newcommand{\pdn}[1]{\nabla\cdot\lp{#1}\rp}
\newcommand{\pdns}[1]{\nabla\lp{#1}\rp}

\newcommand{\vect}[1]{\mathbf{#1}}

\newcommand{\tens}[1]{\mathbf{#1}}

\newcommand{\reff}[1]{Fig.~\ref{#1}}

\newcommand{\reft}[1]{Table~\ref{#1}}

\newcommand{\picsize}{\small}
\newcommand{\picbox}[1]{{#1}}

\journal{ }

\begin{document}

\begin{frontmatter}



\title{Applying Physics-Informed Enhanced Super-Resolution Generative Adversarial Networks to Turbulent Non-Premixed Combustion on Non-Uniform Meshes and Demonstration of an Accelerated Simulation Workflow}


\author[jsc,rwth]{Mathis Bode\corref{cor1}}\ead{m.bode@itv.rwth-aachen.de}
\address[jsc]{{J\"ulich Supercomputing Centre, Forschungszentrum J\"ulich GmbH},{52425 J\"ulich},{Germany}}
\address[rwth]{{Fakult\"at f\"ur Maschinenwesen, RWTH Aachen University},{Templergraben 64},{52056 Aachen},{Germany}}

%

\cortext[cor1]{Corresponding author}

\begin{abstract}
This paper extends the methodology to use physics-informed enhanced super-resolution generative adversarial networks (PIESRGANs) for LES subfilter modeling in turbulent flows with finite-rate chemistry and shows a successful application to a non-premixed temporal jet case. This is an important topic considering the need for more efficient and carbon-neutral energy devices to fight the climate change. Multiple a priori and a posteriori results are presented and discussed. As part of this, the impact of the underlying mesh on the prediction quality is emphasized, and a multi-mesh approach is developed. It is demonstrated how LES based on PIESRGAN can be employed to predict cases at Reynolds numbers which were not used for training. Finally, the amount of data needed for a successful prediction is elaborated.
\end{abstract}



\begin{keyword}
Generative Adversarial Network \sep Direct Numerical Simulation \sep Large-Eddy Simulation \sep Non-Premixed Combustion \sep Simulation Workflows


\end{keyword}

\end{frontmatter}


\section{Introduction}
Artificial intelligence (AI) and Data-driven approaches, such as machine learning (ML) and deep learning (DL), have become very important tools in many scientific domains. This is mainly due to two recent developments. First, ML/DL algorithms have improved drastically with respect to accuracy and robustness, and their execution on graphics processing units (GPUs) became much faster, as the speed of GPUs increased generally and newer GPU generations were developed with a focus on data-driven methods enabling high bandwidths as well as fast and efficient tensor multiplications. Second, the amount of data significantly increased as the size and number of scientific simulations grew, data completeness of experimental setups improved with new techniques, and new data sources appeared, such as mobile phones and the world wide web. Examples for successful applications cover a broad range including speech recognition~\cite{hinton2012deep},  image processing~\cite{dong2014learning,wang2019,greenspan2016guest}, learning of optimal complex control~\cite{Vinyals2019}, and acceleration of drug developments~\cite{bhati2021}. Also, researchers studying turbulence and reactive flows started to use ML/DL tools more frequently. Recent examples are time-advancement for flows by neural networks~\cite{fukami2021nature}, data-driven tabulation as well as optimization of chemistry and combustion~\cite{liang2015,bode2019bspline,DALESSIO2020,chung2021}, DL-supported evaluation of subgrid contributions and probability density functions (PDFs)~\cite{Lapeyre2019,henry2019}, and ML-assisted analysis of combustion behavior~\cite{wan2020}.

%

Among the ML/DL approaches, super resolution is one prominent technique, traditionally developed for image processing applications as single image super-resolution (SISR). The goal was to use data-driven methods to increase the image resolution (i.\,e., to super-resolve the image) by adding information into images to uncover originally hidden features. For this, a network is trained with a large number of images to learn common features. These features are then introduced into the target image based on local information outperforming the prediction accuracy of classical techniques, such as bicubic interpolation.  Dong et al. ~\cite{dong2014learning} relied on a convolutional neural network (CNN) for their super-resolution convolutional neural network~(SRCNN), which learned an end-to-end mapping between the low/high-resolution images.  This approach has been improved over the years \cite{dong2015image,kim2016accurate,kim2016deeply,lai2017deep,tai2017memnet,zhang2018image}. However, as pointed out by Ledig et al.~\cite{ledig2017photo}, these approaches typically produce oversmoothed results, resulting from the attempt to optimize the peak signal-to-noise ratio~(PSNR). Although a high PSNR of a super-resolved image seems to be desirable, it does not necessarily mean that it is visually or perceptually superior.

To improve the high frequency details, which are also of high importance for super-resolution of scientific data, Johnson et al.~\cite{johnson2016perceptual} introduced the concept of perceptual loss, which optimizes the performance regarding the VGG-feature space instead of the PSNR. Furthermore, Ledig et al.~\cite{ledig2017photo} switched to a generative adversarial network (GAN) including residual blocks (RBs), called super-resolution GAN~(SRGAN), instead of a single CNN, in addition to computing the mean squared error (MSE) in the VGG19-feature space~\cite{simonyan2014very} rather than in voxel space. GANs were developed by Goodfellow et al.~\cite{goodfellow2014generative} and aim to learn the unknown data probability distribution of observed data by unsupervised learning, i.\,e., without any labels that are necessary in supervised learning scenarios. They only require access to data samples from the unknown distribution without an explicitly provided data likelihood function. Their training can be also understood as a minimax zero-sum game carried out by two players. The first player, the generator, creates samples that are presented to the second player, the discriminator, which tries to distinguish between generated and real data samples. Both players are coupled by the adversarial loss, and depending on the particular form of this loss term,  finding the equilibrium of this game is equivalent to minimize different distance measures between the generator model and the true data distribution. Examples are Kullback-Leibler (KL) divergence, Jensen-Shannon (JS) divergence, and Wasserstein distance~\cite{arjovsky2017wasserstein}. The SRGAN significantly improved the prediction results with respect to visual similarity and high frequency pattern recovery. However, it produced hallucinated artifacts in the images.  Consequently, Wang et al. ~\cite{wang2018esrgan} developed the enhanced SRGAN~(ESRGAN) to further increase the prediction accuracy. For this, they replaced the RBs by residual in residual dense blocks (RRDBs).

Bode et al.~\cite{bode2019,bode2020dev,bode2021dad} transferred the concept of AI super-resolution to three-dimensional (3-D) turbulence data and developed an AI super-resolution-based subfilter modeling approach for large-eddy simulation (LES) called physics-informed ESRGANs (PIESRGANs, short: PGs). For that, they advanced ESRGANs by introducing a physically informed loss term and enabling it to efficiently handle 3-D data. The general training happens with pairs of high-fidelity data ("H"), such as from fully resolved direct numerical simulations (DNSs), and corresponding filtered data ("F"). During an LES, the trained network generates so-called reconstructed data ("R"), which are used to calculate unclosed subfilter terms by applying a filter kernel and allow to advance the filtered equations accurately in time. PIERSGAN was first applied to turbulence data and spray combustion~\cite{bode2021dad,gauding2021,bode2022dlc,bode2022spray}. More recently, this approach was extended to finite-rate-chemistry flows~\cite{bode2022dpl,bode2022dpt}. A discussion of advantages and shortcoming of this method is available by Bode~\cite{bode2022pdl}.

This work extends PG to non-premixed flows. Moreover, it is demonstrated how non-uniform meshes affect the reconstruction. To correct the resulting numerical errors, a multi-mesh training approach is introduced, which also helps to apply PG more generally in complex simulation scenarios. Finally, the flexibility of AI super-resolution and required amount of training data are discussed. These are crucial developments in the context of LES. Especially, considering the urgency to fight climate change and the resulting need for more efficient and carbon-neutral energy devices, such as turbines powered by hydrogen and engines burning ammonia, accurate subfilter modeling even for complex flows is absolutely mandatory.

This paper is structured as follows. Non-premixed combustion and a non-premixed temporal jet combustion case are outlined in the next section. Afterward, PGs for non-premixed combustion are described. The effect of non-uniform meshes is demonstrated next, and a priori and a posteriori results are presented. A Reynolds number variation and accelerated workflows are subsequently discussed. The paper finishes with conclusions.

\section{Description of non-premixed combustion and a non-premixed temporal jet case}
Fuel and oxidizer are initially separated in non-premixed combustion cases, such as in furnaces as well as diesel and jet engines. The governing equations are given in the next section, and afterward, a non-premixed temporal jet case is briefly described, which is used for analysis in this work.

\subsection{Governing equations}
Simulations of non-premixed combustion are described by conservation equations for mass, species, momentum, and some sort of energy, usually called reactive Navier-Stokes equations (NSEs). Depending on the case sensible specific enthalpy $h$, specific enthalpy including sensible enthalpy and chemical enthalpy $h_\mathrm{sc}$, total specific enthalpy $h_\mathrm{t}$, or the corresponding sensible energy are used. For some low-Mach simulations, a simplified temperature equation is very common. The sensible specific enthalpy is defined as
\begin{equation}
h = \int_{T_0}^{T} \cp \di T,
\end{equation}
where $\cp$ denotes the specific heat capacity at constant pressure and $T$ the temperature with $T_0$ as reference temperature at which the sensible specific enthalpy $h_0$ is defined to be zero. $h_\mathrm{sc}$ is given as
\begin{equation}
h_\mathrm{sc} = h + \sum_{\alpha=1}^{n_\mathrm{s}} \fh Y_\alpha
\end{equation}
with $Y_\alpha$ as mass fractions of the $n_\mathrm{s}$ species and $\fh$ as chemical formation specific enthalpy defined as the enthalpy needed to form \SI{1}{\kilogram} of a species at the reference temperature $T_{\mathrm{f},0}=\SI{298.15}{\kelvin}$. The kinetic energy needs to be added for the total specific enthalpy as
\begin{equation}
h_\mathrm{t} = h_\mathrm{sc}+\dfrac{1}{2} \vect{u}\cdot\vect{u}
\end{equation}
with bold notation for vectors and $\vect{u}$ as velocity vector. Energy and enthalpy are linked by a pressure-based energy term, such as 
\begin{equation}
e_\mathrm{t} = h_\mathrm{t} - \dfrac{p}{\rho}
\end{equation}
for the total sensible energy. Here, $p$ is the pressure, and $\rho$ denotes the density.

Conservation of mass over time $t$ reads
\begin{equation}
\pdt[t]{\rho} + \pdn{\rho \vect{u}} = 0
	\label{dnt:eq:cont}
\end{equation} 
with $\partial_t$ as temporal derivative and the del operator $\nabla$. The mass conservation for species $\alpha$ is
\begin{equation}
	\pdt[t]{\rho Y_\alpha} + \pdn{\rho \lp \vect{u} + \vect{V}_{\mathrm{D},\alpha} \rp Y_\alpha} = \dot{\omega}_\alpha
	\label{dnt:eq:spe}
\end{equation}
with $\vect{V_{\mathrm{D},\alpha}}$ as diffusion velocity vector of species $\alpha$ and $\dot{\omega}_\alpha$ as reaction rate of species $\alpha$. Here, by definition, the identities $\sum_{\alpha=1}^{n_\mathrm{s}} Y_\alpha \vect{V}_{\mathrm{D},\alpha}=\vect{0}$ and $\sum_{\alpha=1}^{n_\mathrm{s}} \dot{\omega}_\alpha=0$ are true. The conservation of momentum is given by
\begin{equation}
	\pdt[t]{\rho \vect{u}} + \pdn{\rho \vect{u}\vect{u}} = -\pdns{p} + \pdn{\tens{\tau}} + \rho \sum_{\alpha=1}^{n_\mathrm{s}} Y_\alpha \vect{f}_\alpha
	\label{dnt:eq:mom}
\end{equation} 
with $\vect{f}_\alpha$ as volume force vector acting on species $\alpha$ and viscous stress tensor $\tens{\tau}$ defined as
\begin{equation}
	\tens{\tau} = \mu \lp( \pdns{\vect{u}} + (\pdns{\vect{u}})^\intercal -\frac{2}{3}\pdn{\vect{u}}\tens{I} \rp
	\label{dnt:eq:visc_stress_tensor}
\end{equation}
with dynamic viscosity $\mu$ and identity tensor $\tens{I}$. As example, the conservation of sensible specific enthalpy gives
\begin{align}
	&& & \pdt[t]{\rho h} + \pdn{\rho \vect{u} h} = -\sum_{\alpha=1}^{n_\mathrm{s}} \fh \dot{\omega}_\alpha + \pdt[t]{p} + \vect{u}\cdot\pdns{p} + \nl +
	 \pdn{\lambda \pdns{T}} - \pdn{\rho \sum_{\alpha=1}^{n_\mathrm{s}} Y_\alpha \vect{V}_{\mathrm{D},\alpha} h_\alpha} + \tens{\tau}\colon\pdns{\vect{u}} + \dot{\mathcal{Q}} + \nl +
	 \rho \sum_{\alpha=1}^{n_\mathrm{s}} Y_\alpha \vect{f}_\alpha \cdot \vect{V}_{\mathrm{D},\alpha}.
	\label{dnt:eq:ent}
\end{align}
Here, $\lambda$ is the thermal conductivity, $h_\alpha$ denotes the sensible specific enthalpy of species $\alpha$, and $\dot{\mathcal{Q}}$ describes a volume source term. This set of equations can be solved by selecting a suitable equation of state (EOS) and model for providing species properties, such as the reaction rates, heat capacities, and diffusion velocity vectors.

\subsection{Case setup of a non-premixed temporal jet}
Denker et al.~\cite{denker2020,denker2021} computed a set of temporally evolving planar non-premixed jets with methane as fuel on a structured mesh denoted by $\vect{\xi}=(\xi_1,\xi_2,\xi_3)^\intercal$ on a domain of size $\overline{L}_1 \times \overline{L}_2 \times \overline{L}_3$. The configuration consists of a fuel stream in the center of a box with two periodic directions ($x_1$- and $x_3$-directions) and outlets in cross-stream direction ($x_2$-direction). The fuel stream moves in positive $x_1$-direction and exhibits initial turbulence. Over time, the area of turbulence expands in y-direction. Additionally, mixing between the fuel and surrounding oxidizer stream happens.

The database features variations with respect to the Reynolds number, defined as 
\begin{equation}
\mathrm{Re} = \frac{\overline{V}_0 H_0}{\nu_\mathrm{fuel}}
\end{equation}
with $\overline{V}_0$ as initial mean jet bulk velocity, $H_0$ as initial jet width, and $\nu_\mathrm{fuel}$ as kinematic viscosity of the fuel. Furthermore, they varied the dilution of the fuel stream, resulting in different stoichiometric mixture fractions $Z_\mathrm{st}$. The scalar dissipation rate, which reads
\begin{equation}
\chi = 2 D \left(\pdns{Z}\right)^2
\end{equation}
with $D$ as diffusion coefficient, quantifies the local mixing in a flow. If the local scalar dissipation rate is too large, the flame quenches. To better understand quenching in non-premixed flames was one of the goals of the dataset by Denker et al.~\cite{denker2020,denker2021}. Therefore, the Damk\"ohler number, given as
\begin{equation}
\mathrm{Da} = \frac{\chi_\mathrm{q} H_0}{V_{\mathrm{c},0}}
\end{equation}
with $\chi_\mathrm{q}=\SI{120}{\per\second}$ as quenching scalar dissipation rate, is another important parameter. Three of their cases are considered in this work: The "Low Re, high dilution case", the "Intermediate Re case", and the "High Re case". As the low dilution case is not used in this work, the three used cases are simply called "LRe", "IRe", and "HRe". The most relevant parameters for the used cases are summarized in Tab.~\ref{dnt:tab:case}. To improve the comparability among the cases, Denker et al.~\cite{denker2020,denker2021} introduced a non-dimensionalized time $t^{*}$. It is shifted by the duration for which the variance of the scalar dissipation rate at stoichiometric mixture faction is zero. Afterward, it is non-dimensionalized by the jet time, defined as $H_0/\overline{V}_0$.
\begin{table}[!htb]
 \centering
 \caption{Summary of parameters for the used non-premixed temporally evolving planar jet cases in this work.}
\label{dnt:tab:case}
\begin{tabular}{c | c c c}
\hline
 & LRe & IRe & HRe \\
 \hline
$\overline{V}_0$ [\si{\meter\per\second}] & \num{16.0} & \num{16.0} & \num{20.7} \\
$H_0$ [\si{\milli\meter}] & \num{18.7} & \num{25.0} & \num{32.3} \\
$\mathrm{Re}$ & \num{4500} & \num{6000} & \num{10000} \\
$\mathrm{Da}$ & \num{0.125} & \num{0.150} & \num{0.150} \\
$Z_\mathrm{st}$ & \num{0.45} & \num{0.45} & \num{0.45} \\
\hline
 \end{tabular}
\end{table}

The cases were computed with the low-Mach solver of the code CIAO, which is also used in this work. CIAO is an arbitrary order finite difference code, which solves the Navier-Stokes equations along with multi-physics effects \cite{desjardins2008}. It is optimized to efficiently run on central processing unit (CPU)-heavy supercomputers~\cite{bode2019hpc,highq15_ciao}. The Poisson equation was solved with the multi-grid solver HYPRE-AMG~\cite{Falgout02_hypre,Henson02_hypre_boomer_amg}, and species as well as temperature equations were discretized with a fifth-order weighted essentially non-oscillatory (WENO) scheme~\cite{Jiang96_weno_scheme}. Furthermore, those equations employed the symmetric operator split of Strang~\cite{Strang68_splitting}. The resulting system of ordinary differential equations for the zero-dimensional homogeneous reactor in each grid cell was solved with a fully time-implicit backward difference method~\cite{Hindmarsh05_sundials,Brown89_cvode}. To simplify the multi-species diffusion, the Hirschfelder and Curtiss approximation~\cite{hirschfelder54_diff_model} was employed with a velocity correction for ensuring species mass conservation. Moreover, the viscous heating was neglected, and changes in the mean pressure were assumed to be small as they are of higher Mach number order. Volume forces and volume source term did not occur.  Consequently, a simplified set of momentum, species and temperature equations was solved. It reads
\begin{equation}
	\pdt[t]{\rho Y_\alpha} + \pdn{\rho \lp \vect{u} + \vect{V}_{\mathrm{H},\alpha} \rp Y_\alpha} = \pdn{\rho D_\alpha \frac{W_\alpha}{W} \pdns{X_\alpha}} + \dot{\omega}_\alpha
	\label{dnt:eq:sspe}
\end{equation}
for species mass conservation with $\vect{V}_{\mathrm{H},\alpha}=\sum_{\alpha=1}^{n_\mathrm{s}} D_\alpha \frac{W_\alpha}{W}\pdns{X_\alpha}$ as correction velocity ensuring species mass conservation after applying the Hirschfelder and Curtiss approximation with $W$ as mean molecular weight, which is computed as $\sum_{\alpha=1}^{n_\mathrm{s}} X_\alpha W_\alpha$ with $X_\alpha$ and $W_\alpha$ being the mole fraction and molecular weight of species $\alpha$. $D_\alpha$ is the diffusion coefficient, which can be computed from binary mass diffusion coefficients as 
\begin{equation}
D_\alpha = \frac{1-Y_\alpha}{\sum_{\alpha\neq\beta}X_\beta/D_{\beta\alpha}}.
\end{equation}
The conservation of momentum is given by
\begin{equation}
	\pdt[t]{\rho \vect{u}} + \pdn{\rho \vect{u}\vect{u}} = -\pdns{p} + \pdn{\tens{\tau}}.
	\label{dnt:eq:smom}
\end{equation} 
The temperature equation results in
\begin{align}
	&& & \pdt[t]{\rho \cp T} + \pdn{\rho \cp \vect{u} T} = -\sum_{\alpha=1}^{n_\mathrm{s}} \lp h + \fh \rp \dot{\omega}_\alpha + \nl + \pdn{\lambda \pdns{T}} - \rho \pdns{T} \cdot \pdn{\sum_{\alpha=1}^{n_\mathrm{s}} \cp D_\alpha \frac{W_\alpha}{W} \pdns{X_\alpha}}.
	\label{dnt:eq:sent}
\end{align}

The used mechanism features 28 species and 102 reactions \cite{peters2002} for the oxidation of methane and is complemented by the Zeldovich mechanism for NO formation \cite{lavoie1970}. To save computing time and increase the numerical dissipation towards the cross-stream outlets, a non-uniform mesh was employed with increasing cell width in cross-stream direction.

\section{PIESRGAN for turbulent non-premixed combustion} 
Rather than solving the fully resolved reactive NSEs as in DNSs, LESs solve a filtered form of the NSEs. Generally, a homogeneous filter operation, which furthermore is assumed to be a Reynolds operator and denoted with an overbar, can be used to define the Reynolds decomposition as
\begin{equation}
  \label{dtr:eq:split}
  \{\cdot\} = \rfilt{\{\cdot\}} + \rfluc{\{\cdot\}}
\end{equation}
with the overbar denoting the filtered part and the prime the subfilter part of a quantity with 
\begin{equation}
  \label{dtr:eq:id1}
  \rfilt{\rfilt{\{\cdot\}}} = \rfilt{\{\cdot\}}\rfilt{1} = \rfilt{\{\cdot\}}
\end{equation}
and
\begin{equation}
  \label{dtr:eq:id2}
  \rfilt{\rfluc{\{\cdot\}}} = 0.
\end{equation}
Additionally, for flows with variable density, such as reactive flows, the Favre-filtering is commonly defined as 
\begin{equation}
 \filt{\{\cdot\}} = \frac{\rfilt{\{\rho \cdot\}}}{\rfilt{\rho}}.
\end{equation}
The corresponding Favre decomposition reads
\begin{equation}
\{\cdot\} = \filt{\{\cdot\}} + \fluc{\{\cdot\}}.
\end{equation}
Here, the identity $\rfilt{\fluc{\rho\{\cdot\}}} = 0$ hold but $\rfilt{\fluc{\{\cdot\}}} \neq 0$.

For the reactive LES equations, Favre decomposition is introduced in the reactive NSEs, and a Reynolds filter is applied to all terms. This leads to 
\begin{equation}
\pdt[t]{\rfilt{\rho}} + \pdn{\rfilt{\rho}\filt{\vect{u}}} = 0
	\label{dnt:eq:fcont}
\end{equation} 
for the mass conservation,
\begin{equation}
	\pdt[t]{\rfilt{\rho} \filt{Y}_\alpha} + \pdn{\rfilt{\rho} \filt{\vect{u}} \filt{Y}_\alpha} = \pdn{\rfilt{\rho \vect{V}_{\mathrm{D},\alpha} Y_\alpha} - \rfilt{\fluc{\vect{u}}\fluc{Y_\alpha}}} + \rfilt{\dot{\omega}_\alpha}
	\label{dnt:eq:fspe}
\end{equation}
for the species mass conservation,
\begin{equation}
	\pdt[t]{\rfilt{\rho} \filt{\vect{u}}} + \pdn{\rfilt{\rho} \filt{\vect{u}}\filt{\vect{u}}} = -\pdns{\rfilt{p}} + \pdn{\rfilt{\tens{\tau}} - \rfilt{\rho} \rfilt{\fluc{\vect{u}}\fluc{\vect{u}}}} + \rfilt{\rho \sum_{\alpha=1}^{n_\mathrm{s}} Y_\alpha \vect{f}_\alpha}
	\label{dnt:eq:fmom}
\end{equation}
for the momentum conservation, and
\begin{align}
	&& & \pdt[t]{\rfilt{\rho} \filt{h}} + \pdn{\rfilt{\rho} \filt{\vect{u}} \filt{h}} =  \pdt[t]{\rfilt{p}} + \rfilt{\vect{u}\cdot\pdns{p}} + \nl +
	 \pdn{\rfilt{\lambda \pdns{T}} - \rfilt{\rho}\rfilt{\fluc{\vect{u}}\fluc{h}}} - \pdn{\rfilt{\rho \sum_{\alpha=1}^{n_\mathrm{s}} Y_\alpha \vect{V}_{\mathrm{D},\alpha} h_\alpha}} + \nl + \rfilt{\tens{\tau}\colon\pdns{\vect{u}}} + \rfilt{\dot{\mathcal{Q}}} +
	 \rfilt{\rho \sum_{\alpha=1}^{n_\mathrm{s}} Y_\alpha \vect{f}_\alpha \cdot \vect{V}_{\mathrm{D},\alpha}}
	\label{dnt:eq:fent}
\end{align}
for the conservation of the sensible specific enthalpy.

PG-based subfilter models can be used to systematically close all terms in the filtered NSEs. The algorithm and details are discussed next.

\subsection{Loss function}
The loss function is the target function of the problem which is minimized by the optimization solver. PG extends traditional loss functions by a physically motivated term, which prioritizes the fulfillment of important flow conditions, such as mass and species conservation, in the generated data. For example, a violation of the mass conservation condition usually leads to blowing up simulations and therefore the described weak enforcement is required.  The loss function reads
\begin{equation}
\mathcal{L} = \beta_1 L_\mathrm{adversarial} + \beta_2 L_\mathrm{pixel} + \beta_3 L_\mathrm{gradient} + \beta_4 L_\mathrm{physics},
\label{dnt:eq:loss}
\end{equation}
where $\beta_1$, $\beta_2$, $\beta_3$, and $\beta_4$ are coefficients weighting the different loss term contributions. $L_\mathrm{adversarial}$ is the discriminator/generator relativistic adversarial loss \cite{jolicoeur2018relativistic}. It measures both how well the generator is able to create data and how well the discriminator is able to identify fake data. $L_\mathrm{pixel}$ is defined using the mean-squared error (MSE) of the quantity.  $L_\mathrm{gradient}$ computes the MSE of the gradient of a quantity.  The physical loss term is
\begin{equation}
L_\mathrm{physics} = \beta_{41} L_\mathrm{mass} + \beta_{42} L_\mathrm{species} + \beta_{43} L_\mathrm{elements},
\label{dnt:eq:pl}
\end{equation}
where $\beta_{41}$, $\beta_{42}$, and $\beta_{43}$ are coefficients weighting the different physical loss term contributions. It accounts for mass conservation, species conservation, and elements conservation. Bode~\cite{bode2022dpl} and Bode et al.~\cite{bode2022dpt} pointed out that elements conservation was not required for their application cases. It turned out that this is dependent on the chemical reaction mechanism and the combustion regime. For the mechanism employed in this work, the elements loss term improved the prediction accuracy considerably. Both sets of weighting coefficient sum up to unity as $\sum_{i} \beta_i = 1$ and $\sum_{i} \beta_{4i} = 1$.

\subsection{Architecture} 
\begin{figure*}[htbp]
\centerline{\includegraphics[width=\textwidth]{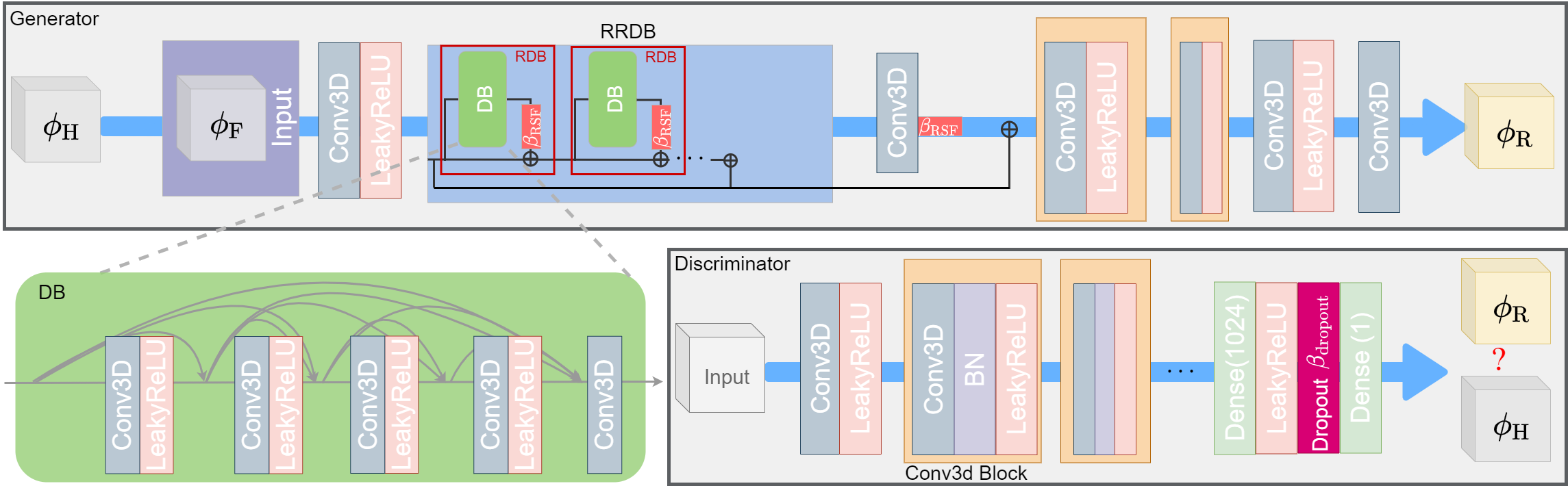}}
\caption{Sketch of PG. "H" denotes high-fidelity data, such as DNS data, "F" are corresponding filtered data, and "R" are the reconstructed data.  The components are: Conv3D - 3D Convolutional Layer, LeakyReLU - Activation Function, DB - Dense Block, RDB - Residual Dense Block, RRDB - Residual in Residual Dense Block, $\beta_\mathrm{RSF}$ - Residual Scaling Factor, BN - Batch Normalization, Dense - Fully Connected Layer, Dropout - Regularization Component, $\beta_\mathrm{dropout}$ - Dropout Factor. Image from \cite{bode2021dad}.}
\label{dnt:fig:piesrgan}
\end{figure*}

The PG architecture is depicted in \reff{dnt:fig:piesrgan}. Training is done with pairs of data ($\phi_\mathrm{H}$ and $\phi_\mathrm{F}$), which are computed in a prestep if training is done with DNS data. The generator creates the reconstructed data $\phi_\mathrm{R}$ with $\phi_\mathrm{F}$ as input. $\phi_\mathrm{H}$ is used to evaluate all loss function terms. Both generator and discriminator use 3-D CNN layers (Conv3D) \cite{krizhevsky2012imagenet} with kernel size of 3 and stride 1 combined with leaky rectified linear unit (LeakyReLU) layers for activation \cite{maas2013}. Furthermore, the generator features a residual in residual dense block (RRDB), which contains fundamental architectural elements such as residual dense blocks (RDBs) with skip-connections, an extended residual block (RB) with dense connections inside. Also, residual scaling factors $\beta_{\mathrm{RSF}}$ are included in order to avoid instabilities in the forward and backward propagation. On the other hand, the discriminator has additional layers for batch normalization (BN) as well as a dropout with dropout factor $\beta_{\mathrm{dropout}}$ and a final dense layer (Dense). In total, the generator has 80 layers, while the discriminator has only 28 layers for all cases considered in this work.

\subsection{Hyperparameters} 
The hyperparameters are summarized in Tab.~\ref{dnt:tab:hp}, which were optimized as described by Bode~\cite{bode2022dpl} in the context of species splitting. In addition to loss functions focussing on single time steps, multi-time step terms were employed. Generally, it was found that the network gives good results in a wider range of hyperparameters. Furthermore, a working hyperparameter set usually also gives good results for other related cases.
\begin{table}[!htb]
 \centering
 \caption{Overview of the PG hyperparameters. The given ranges represent the sensitivity intervals with acceptable network results. The central values were used in this work. Note that the given boundaries are not combined sets of hyperparameters and therefore do not sum up to one.}
\label{dnt:tab:hp}
\begin{tabular}{c c}
\hline
$\beta_1$ & $[\num{0.2e-5},\num{0.6e-4},\num{0.8e-4}]$ \\
$\beta_2$ & $[\num{0.79327},\num{0.83994},\num{0.91812}]$ \\
$\beta_3$ & $[\num{0.04},\num{0.06},\num{0.15}]$ \\
$\beta_4$ & $[\num{0.04},\num{0.10},\num{0.15}]$ \\
$\beta_{41}$ & $[\num{0.30},\num{0.60},\num{0.85}]$ \\
$\beta_{42}$ & $[\num{0.20},\num{0.30},\num{0.45}]$ \\
$\beta_{43}$ & $[\num{0.05},\num{0.10},\num{0.15}]$ \\
$\beta_\mathrm{RSF}$ & $[\num{0.1},\num{0.2},\num{0.3}]$ \\
$\beta_\mathrm{dropout}$ & $[\num{0.2},\num{0.4},\num{0.5}]$ \\
$l_\mathrm{generator}$ & $[\num{1.2e-6},\num{4.5e-6},\num{5.0e-6}]$ \\
$l_\mathrm{discriminator}$ & $[\num{4.4e-6},\num{4.5e-6},\num{8.5e-6}]$ \\
\hline
 \end{tabular}
\end{table}

\subsection{Species splitting}
Species splitting as introduced by Bode~\cite{bode2022dpl} was used for the production runs in this work. Consequently, the set of species was split into primary and secondary species. While the secondary species were updated at the same time as the velocities, the update of the primary species is shifted by a half time step. Furthermore, the prediction accuracy for the primary species is improved by solving an additional transport equation on the reconstructed mesh. \reft{dnt:tab:species} lists the maximum mass fraction of each species at the later time of the HRe case to give an impression of the impact of each species on quantities such as the mixture fraction. The AutoML optimization for the case considered in this work resulted in $\mathrm{N}_2$, $\mathrm{CH}_4$, $\mathrm{OH}$, $\mathrm{H}_2$, $\mathrm{CO}$, $\mathrm{CO}_2$, and $\mathrm{C}_2\mathrm{H}_3$ as set of primary species.
\begin{table}[!htb]
 \centering
 \caption{Overview of the maximum mass fraction of each species at the later time for the HRe case.}
\label{dnt:tab:species}
\begin{tabular}{c c}
\hline
Species name $\alpha$ & Maximum mass fraction $\max \lp Y_{\alpha}^{\mathrm{late}} \rp$ \\
\hline
$\mathrm{N}_2$ & \num[scientific-notation=true,round-mode=places,round-precision=4]{0.92385147561885} \\
$\mathrm{O}_2$ & \num[scientific-notation=true,round-mode=places,round-precision=4]{0.232181222116047} \\
$\mathrm{H}$ & \num[scientific-notation=true,round-mode=places,round-precision=4]{0.000230226856251529} \\
$\mathrm{OH}$ & \num[scientific-notation=true,round-mode=places,round-precision=4]{0.0023830686568589} \\
$\mathrm{O}$ & \num[scientific-notation=true,round-mode=places,round-precision=4]{0.00178991714836124} \\
$\mathrm{H}_2$ & \num[scientific-notation=true,round-mode=places,round-precision=4]{0.00154416246233481} \\
$\mathrm{H}_2\mathrm{O}$ & \num[scientific-notation=true,round-mode=places,round-precision=4]{0.0737483162689063} \\
$\mathrm{HO}_2$ & \num[scientific-notation=true,round-mode=places,round-precision=4]{0.000132267800003505} \\
$\mathrm{H}_2\mathrm{O}_2$ & \num[scientific-notation=true,round-mode=places,round-precision=4]{2.50660516662027e-05} \\
$\mathrm{CO}$ & \num[scientific-notation=true,round-mode=places,round-precision=4]{0.0312029629711663} \\
$\mathrm{CO}_2$ & \num[scientific-notation=true,round-mode=places,round-precision=4]{0.0898053152891907} \\
$\mathrm{CH}$ & \num[scientific-notation=true,round-mode=places,round-precision=4]{7.22143416499925e-06} \\
$\mathrm{CHO}$ & \num[scientific-notation=true,round-mode=places,round-precision=4]{1.26155806561309e-05} \\
$\mathrm{CH}_2$ & \num[scientific-notation=true,round-mode=places,round-precision=4]{1.45344231706473e-05} \\
$\mathrm{CH}_2\mathrm{O}$ & \num[scientific-notation=true,round-mode=places,round-precision=4]{0.000955273014587293} \\
$\mathrm{CH}_3$ & \num[scientific-notation=true,round-mode=places,round-precision=4]{0.00101961659511231} \\
$\mathrm{CH}_4$ & \num[scientific-notation=true,round-mode=places,round-precision=4]{0.0647854244686873} \\
$\mathrm{C}_2\mathrm{H}_6$ & \num[scientific-notation=true,round-mode=places,round-precision=4]{0.00163486277317824} \\
$\mathrm{C}_2\mathrm{H}$ & \num[scientific-notation=true,round-mode=places,round-precision=4]{5.15543351026059e-07} \\
$\mathrm{C}_2\mathrm{H}_2$ & \num[scientific-notation=true,round-mode=places,round-precision=4]{0.00331811069529786} \\
$\mathrm{CHCO}$ & \num[scientific-notation=true,round-mode=places,round-precision=4]{0.000678054977242168} \\
$\mathrm{C}_3\mathrm{H}_3$ & \num[scientific-notation=true,round-mode=places,round-precision=4]{5.93284435033753e-05} \\
$\mathrm{C}_2\mathrm{H}_3$ & \num[scientific-notation=true,round-mode=places,round-precision=4]{2.37248724125937e-05} \\
$\mathrm{C}_2\mathrm{H}_4$ & \num[scientific-notation=true,round-mode=places,round-precision=4]{0.00137748081910278} \\
$\mathrm{C}_2\mathrm{H}_5$ & \num[scientific-notation=true,round-mode=places,round-precision=4]{8.23456305856425e-05} \\
$\mathrm{C}_3\mathrm{H}_4$ & \num[scientific-notation=true,round-mode=places,round-precision=4]{0.000661999533990321} \\
$\mathrm{C}_3\mathrm{H}_5$ & \num[scientific-notation=true,round-mode=places,round-precision=4]{0.000248801664833752} \\
$\mathrm{C}_3\mathrm{H}_6$ & \num[scientific-notation=true,round-mode=places,round-precision=4]{0.000906313979534389} \\
$\mathrm{N}$ & \num[scientific-notation=true,round-mode=places,round-precision=4]{2.91274526773288e-10} \\
$\mathrm{NO}$ & \num[scientific-notation=true,round-mode=places,round-precision=4]{3.48587339521417e-06} \\
\hline
 \end{tabular}
\end{table}

\subsection{Algorithm} 
Once the PG is trained, it is used to reconstruct all required data fields during every time step of a LES. This is part of a multi-step algorithm which computes all unclosed terms in the LES equations systematically. The algorithm starts with the LES solution $\Phi_\mathrm{F}^{n}$ at time step $n$, which includes the entirety of all fields in the simulation except the primary species, and the LES solution of the primary species $\Xi_\mathrm{F}^{n+1/2}$ at time step $n+1/2$. It consists of repeating the following steps:
\begin{enumerate}
\item Use the PG to reconstruct $\Phi_\mathrm{R}^n$ from $\Phi_\mathrm{LES}^n$ with  $\Xi_\mathrm{F}^{n+1/2}$ as additional information.
\item Use $\Phi_\mathrm{R}^{n}$ to estimate the unclosed terms $\Psi_\mathrm{LES}^n$ in the LES equations for all $\Phi$ fields by applying a filter operator.
\item Use $\Psi_\mathrm{LES}^n$ and $\Phi_\mathrm{LES}^n$ to advance the LES equations of $\Phi$ to $\Phi_\mathrm{LES}^{n+1}$.
\item Use the PG to reconstruct $\Xi_\mathrm{R}^{n+1/2}$ from $\Xi_\mathrm{LES}^{n+1/2}$.
\item Use $\Xi_\mathrm{R}^{n+1/2}$ to update the fields of $\Xi$ to $\Xi_\mathrm{R}^{n+1/2;\mathrm{update}}$ by solving the unfiltered scalar equations on the mesh of $\Xi_\mathrm{R}^{n+1/2}$ with $\Phi_\mathrm{R}^{n}$ as additional information.
\item Use $\Xi_\mathrm{R}^{n+1/2;\mathrm{update}}$ to estimate the unclosed terms $\Gamma_\mathrm{LES}^{n+1/2}$ in the LES equations of $\Xi$ for all fields by evaluating the local terms with $\Xi_\mathrm{R}^{n+1/2;\mathrm{update}}$, $\Phi_\mathrm{F}^{n}$, $\Phi_\mathrm{F}^{n+1}$, and $\Phi_\mathrm{R}^{n}$ as well as applying a filter operator.
\item Use $\Gamma_\mathrm{LES}^{n+1/2}$ and $\Xi_\mathrm{LES}^{n+1/2}$ to advance the LES equations of $\Xi$ to $\Xi_\mathrm{LES}^{n+3/2}$.
\end{enumerate}

\subsection{Training process} 
\label{dnt:ssec:training}
The RMSProp solver, relying on the stochastic gradient descent (SGD) approach, was used as optimizer during training in this work. Stability issues are the main problem during the training process of complex networks and especially GANs. In addition to standard difficulties in training deep neural networks which are due to traversing highly complex landscapes of the loss function spanned by the network, the intricate interplay of the two components of GANs, discriminator and generator, leads to further vulnerabilities during the training process. Standard techniques, such as adaptive learning rates, were employed in this work. Furthermore, it was found that normalization is crucial for the training success. Turbulent fluctuating fields were zero mean-centered and rescaled with their root-mean-square deviation (RMSD) value. The magnitude of the mass fraction fields were aligned with an estimate for the maximum value, mapping all fields between zero and about one.

The training with multiple cases, i.e., LRe and HRe for the application case in Sec.~\ref{dnt:ssec:rey}, is particularly challenging and sometimes, the loss function does not decrease further, as a local (non-optimum) minimum was found. Slightly disturbing the beta coefficients in Eqs.~\ref{dnt:eq:loss} and \ref{dnt:eq:pl} helped to overcome this lock and further improve the solution.

Generally, the networks were initialized with the solution for decaying turbulence for varying Reynolds numbers from Bode et al.~\cite{bode2021dad}. One way to improve the stability during the training process is to only update the generator network weights afterward. This makes it more difficult for the network to deviate from the solution of fully homogeneous isotropic turbulence (HIT), which was used for the decaying turbulence case. In this view, the discriminator supervises that the solution remains close to a turbulent solution, which might be also a problem for some cases, but gave good results in this work. An advantage of this approach is that one degree of freedom for the training process is eliminated, making the training more handy. 

Furthermore, it was found that it is essential to use consistent (high-order) numerics during the training process, such as for evaluating the loss function terms. Otherwise small numerical errors accumulate during the training process leading to deviations compared to the DNS solution. One way to check the quality of the achieved results is by means of the energy spectrum. If the numerics are not consistent and accurate enough, deviations for high wavenumbers are inevitable.


\subsection{Runtime execution} 
\label{dnt:ssec:runtime}
On runtime, the same normalization as during training should be used. In order to efficiently execute the computations, reconstruction of data fields and consecutive filtering should be done on GPUs.


\subsection{Implementation details} 
A PG implemented as part of a TensorFlow/Keras framework~\cite{abadi2015tensorflow,chollet2015keras} was used in this work. It was coupled to the simulation code CIAO employing the TensorFlow C API. During training and execution, subboxes were considered to reduce the memory requirements. A subbox size of $16\times16\times16$ gave good results. In general, larger subboxes increase the accuracy, however, also super-linearly increase the computing cost.

Training can be done with stored data or on-the-fly along with a DNS. On-the-fly training is more efficient as storing cost can be reduced, which are usually non-negligible for large cases. Especially, if transient processes need to be learnt, the requirements with respect to the temporal resolution can be problematic.  Additionally, on-the-fly training can be a smart usage of GPUs for simulation codes which cannot natively make use of GPUs, but need to run on computing clusters where all nodes are equipped with GPUs. Moreover, on-the-fly training can improve the convergence of the training process, as the transition is "quasi-continuous" and not discrete due to the storing frequency. An issue with respect to on-the-fly training is that data are only available once, which makes iterative update processes difficult.

A base implementation of PG can be found on GitLab (https://git.rwth-aachen.de/Mathis.Bode/PIESRGAN.git), and a detailed discussion of computational aspects is given by Bode~\cite{bode2022sra}.

\section{PIESRGAN for non-uniform meshes} 
Bode et al.~\cite{bode2021dad} did not discuss the effect of non-uniform meshes on PG-subfilter modeling. However, it can be expected that the model is mesh dependent, especially if a mesh-dependent filter is employed, whose filter width implicitly is subject to the local cell size.  The mesh is also critical during up- and downsampling, as the LES has typically a much coarser mesh than the DNS. Obviously, it is not possible to do the reconstruction on the LES mesh, as it is not able to resolve the added information.

The effect can be discussed by introducing multiple meshes and mesh operators: The LES mesh $D_\mathrm{LES}$, the DNS mesh $D_\mathrm{DNS}$, the mesh for the reconstruction $D_\mathrm{R}$,  the mesh after filtering the DNS data $D_\mathrm{F}$, the mesh after filtering the reconstructed data $D_\mathrm{RF}$, the mesh operator to interpolate the values on the LES mesh to the DNS mesh ($D_\mathrm{LES} \rightarrow D_\mathrm{DNS}$) $\mathscr{I}_{\mathrm{LES}\rightarrow\mathrm{DNS}}$, the mesh operator to filter the data discretized on the mesh for the reconstruction, evaluate the unclosed terms, and store the resulting field on a mesh for filtering ($D_\mathrm{R} \rightarrow D_\mathrm{RF}$) $\mathscr{F}_{\mathrm{R}\rightarrow\mathrm{RF}}$, the mesh operator to interpolate the values on the mesh for filtering to the LES mesh ($D_\mathrm{RF} \rightarrow D_\mathrm{LES}$) $\mathscr{I}_{\mathrm{RF}\rightarrow\mathrm{LES}}$, and the mesh operator to filter the data discretized on the DNS mesh and store the resulting field on a mesh for filtering ($D_\mathrm{DNS} \rightarrow D_\mathrm{F}$) $\mathscr{F}_{\mathrm{DNS}\rightarrow\mathrm{F}}$. Bode et al.~\cite{bode2021dad} used a PG to reconstruct data discretized on the DNS mesh on the mesh for reconstruction, denoted as $\mathscr{P}_{\mathrm{DNS}\rightarrow\mathrm{R}}$. 

Using these notations, computing data pairs for training can be expressed as
\begin{equation}
\mathscr{F}_{\mathrm{DNS}\rightarrow\mathrm{F}} \circ D_\mathrm{DNS},
\end{equation}
which maps $D_\mathrm{LES}\rightarrow D_\mathrm{F}$. The trivial implementation used for execution is
\begin{equation}
\mathscr{I}_{\mathrm{RF}\rightarrow\mathrm{LES}} \circ \mathscr{F}_{\mathrm{R}\rightarrow\mathrm{RF}} \circ \mathscr{P}_{\mathrm{DNS}\rightarrow\mathrm{R}} \circ \mathscr{I}_{\mathrm{LES}\rightarrow\mathrm{DNS}} \circ D_\mathrm{LES}.
\label{dnt:eq:simp}
\end{equation}
This results in $D_\mathrm{LES} \rightarrow D_\mathrm{DNS} \rightarrow D_\mathrm{R} \rightarrow D_\mathrm{RF} \rightarrow D_\mathrm{LES}$. 

Bode et al.~\cite{bode2021dad} chose $D_\mathrm{F}=D_\mathrm{R}=D_\mathrm{DNS}$, as their DNS data used a uniform mesh with cubes and in order to support the evaluation of the loss function terms. Furthermore,  they used $D_\mathrm{RF}=D_\mathrm{LES}$, which is only correct if the LES uses a uniform mesh. This reduces the number of different meshes involved but has three major limitations: First, it only works for uniform meshes, as the effect of the non-uniform mesh on the filter is not considered. Second, the DNS mesh might be very large and thus using the DNS mesh as basis mesh can be difficult from a memory usage point of view. Finally, the additional interpolation operators, which need to be employed on runtime, can be expensive and introduce further numerical errors. Note that subboxes of the domain are used for training and reconstruction, i.\,e., the mesh resolutions must match, however, the minimum and maximum global values of the meshes do not need to match, and full meshes are assembled by subboxes. As described earlier, this is important for computational reasons, but also essential for making the network generally employable for different cases.

To overcome the limitations with respect to non-uniform meshes, a new computational domain $\Omega$ is defined as base mesh with the coordinates $\vect{\xi}=(\xi_1,\xi_2,\xi_3)^\intercal$. $\Omega$ is a rectangular algebraic mesh with uniform grid spacing in all directions. This is in contrast to the other defined meshes, which can be any kind of mesh, as long as a uni-directional mapping between $D$ (with coordinates $\vect{x}=(x_1,x_2,x_3)^\intercal$) and $\Omega$ exist, i.\,e., the transformation can be done by a simple tensor operation. The resolution of $\Omega$ should be chosen similar to finest resolution in the DNS mesh. To account for the different meshes during training, the Jacobian is introduced as
\begin{equation}
J=\frac{\delta x}{\delta \xi}
\end{equation}
and computed in every mesh cell. The Jacobian is used as additional input field for the PG reconstruction, which reconstructs from $D_\mathrm{LES}$ to $\Omega$, denoted as $\mathscr{P}_{\mathrm{LES}\rightarrow\Omega}$.  Furthermore, the mesh operator to filter the data discretized on the computational domain, evaluate the unclosed terms, and store the resulting field on the LES mesh ($\Omega \rightarrow D_\mathrm{LES}$) $\mathscr{F}_{\Omega\rightarrow\mathrm{LES}}$ is introduced. The execution process becomes
\begin{equation}
\mathscr{F}_{\Omega\rightarrow\mathrm{LES}} \circ \mathscr{P}_{\mathrm{LES}\rightarrow\Omega} \circ D_\mathrm{LES}.
\label{dnt:eq:corr}
\end{equation}
The additional input field $J$ is used by the network to consider the effect of the local mesh resolution. As it is impossible to train the network with all relevant $J$ values, if the LES changes its local mesh size continuously or adaptively, the network needs to be enabled to do some kind of interpolation with respect to the filter width. For some particular filter kernels, this can be done analytically. However, it is more flexible to integrate this interpolation in the GAN. Therefore, the GAN is always trained with two data triples with different $J$, corresponding to different mesh resolution of $\Omega$, in this work. As will be shown in the next section, this significantly improves the results on non-uniform meshes, such as the target non-premixed temporal jet case in this work. However, this also increases the training cost, as multiple meshes need to be considered.

The described approach for non-uniform meshes has another advantage. If multiple datasets with different resolutions, such as turbulence data with different Reynolds numbers, are used for training, multiple mesh resolutions can be used for training without initial interpolation, employing the Jacobian approach.

\subsection{Effect of non-uniform meshes} 
\begin{figure}[!tb]
\picsize
	\centering
    \picbox{\input{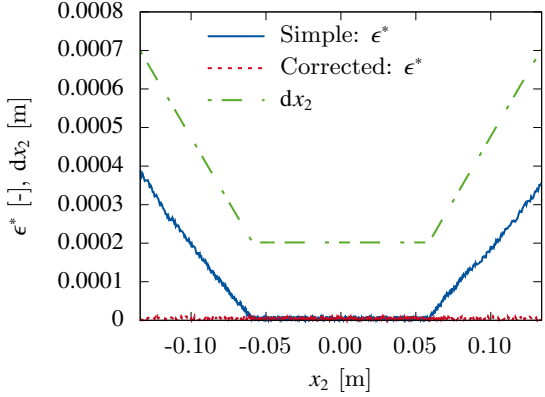}}
	\caption{Plot of the nondimensional error $\epsilon^{*}$ as function of the $x_2$-coordinate. Furthermore, the cell size in $x_2$-direction $\mathrm{d}x_2$ is shown.}
	\label{dnt:fig:gerr}
\end{figure}

The effect of the non-uniform mesh is discussed by means of the HRe case, which features uniform meshes in the $x_1$- and the $x_3$-direction. The mesh in $x_2$-direction is uniform in the center but the mesh increment increases towards the boundaries. A relative error based on the velocity field is defined as
\begin{equation}
\epsilon^{*}=\frac{\overline{L}_2 \int_{\overline{L}_1} \int_{\overline{L}_3} \sqrt{(\vect{u}_{\mathrm{R}}-\vect{u}_{\mathrm{H}})\cdot(\vect{u}_{\mathrm{R}}-\vect{u}_{\mathrm{H}})} \di x_3 \di x_1}{\int_{\overline{L}_1} \int_{\overline{L}_2} \int_{\overline{L}_3} \sqrt{\vect{u}_{\mathrm{H}}\cdot\vect{u}_{\mathrm{H}}} \di x_3 \di x_2 \di x_1}.
\end{equation}
Figure~\ref{dnt:fig:gerr} shows the relative errors plotted as function of $x_2$ for the simple algorithm (cf.~Eq.~\ref{dnt:eq:simp}) and the corrected algorithm with Jacobian as input (cf.~Eq.~\ref{dnt:eq:corr}). Furthermore, the mesh increment in $x_2$-direction is plotted. While the corrected algorithm gives accurate results, the simple algorithm shows deviations which seem to correlate with the mesh increment in $x_2$-direction. As the finer resolution in the center was used for training with the simple algorithm, the network seems to add the wrong amount of information in the stretched areas. The result is remarkable for two reasons: First, the accuracy of the corrected algorithm over the full width is very good. Second, the error introduced in the stretched regions by the simple algorithm is surprising, as the flow field is very weak there and not turbulent.

\subsection{A priori testing}
To evaluate the prediction accuracy of PG for the non-premixed jet case, a priori results are discussed by means of cases LRe and HRe. A separate network was used for each case. The networks were trained with multiple time steps from the case. To have a weak distinction between training and test data, the time step used for testing was not used for training. This is not optimal, however, as only one realization per case exists, further distinction between training and test data was not possible. For the cases presented in the application section, a better distinction between training and testing data was ensured.

The analysis is done by means of mass fraction fields, velocity component fields, mixture fraction fields, and scalar dissipation rate fields. The time staggering is a double problem here: First, not all mass fractions are stored at the same time, as they are shifted by a half time step due to the species splitting. This was not done in the original DNS, which requires the original DNS data for these mass fractions to be interpolated in time using two consecutive time steps. Second, the mixture fraction and consecutively scalar dissipation rate are not solved quantities but computed by means of the solved mass fractions. Therefore, the mass fractions need to be interpolated to the same time step. As the primary species are the dominant species, the secondary species were interpolated by a half time step, using two reconstructed data fields, which required three DNS time steps. 

\begin{figure}[!htb]
    \picsize
    \centering
    \begin{subfigure}[b]{\columnwidth}
    \centering
        \begin{subfigure}[b]{22mm}
            \centering
            \includegraphics[width=\textwidth]{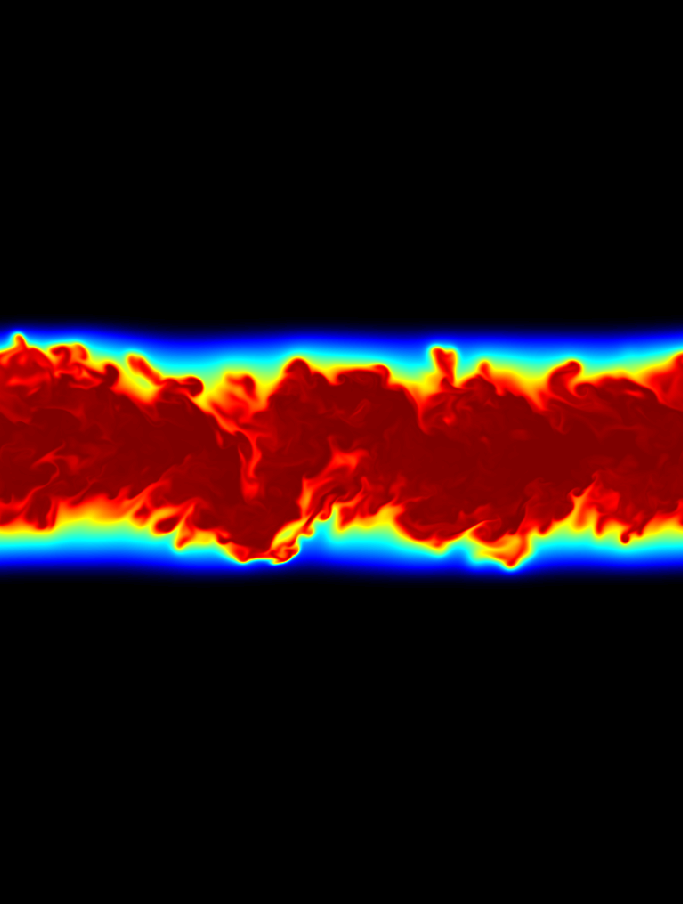}
            \vspace*{-10mm}
            \caption*{\colorbox{white}{$Z_\mathrm{H}$}}  
            \label{dnt:sfig:over4e:1H}
        \end{subfigure}
        \hspace{0mm}
        \begin{subfigure}[b]{22mm}
            \centering
            \includegraphics[width=\textwidth]{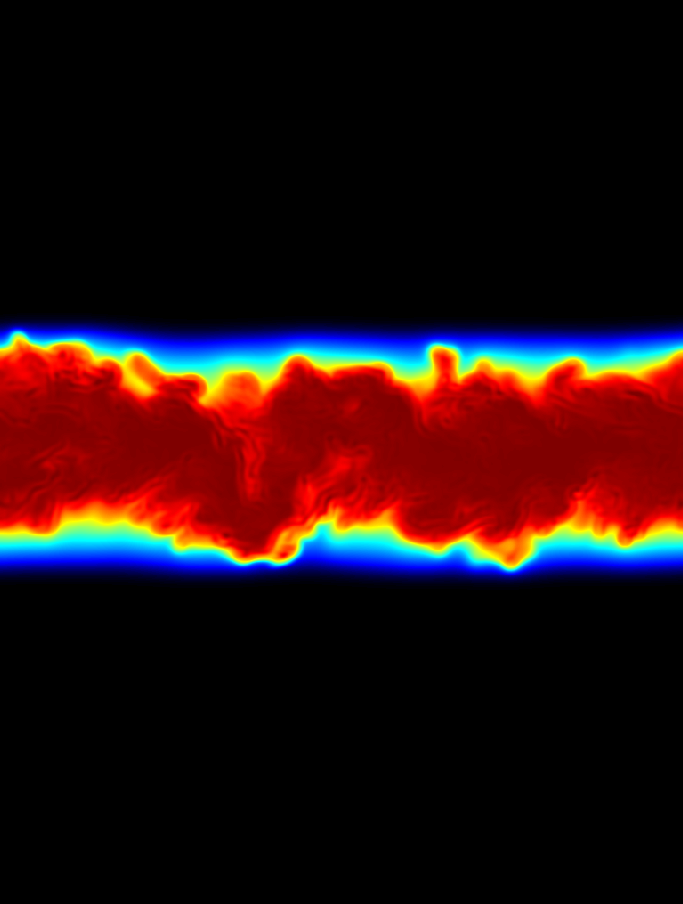}
            \vspace*{-10mm}
            \caption*{\colorbox{white}{$Z_\mathrm{F}$}}  
            \label{dnt:sfig:over4e:1F}
        \end{subfigure}
        \hspace{0mm}
        \begin{subfigure}[b]{22mm}
            \centering
            \includegraphics[width=\textwidth]{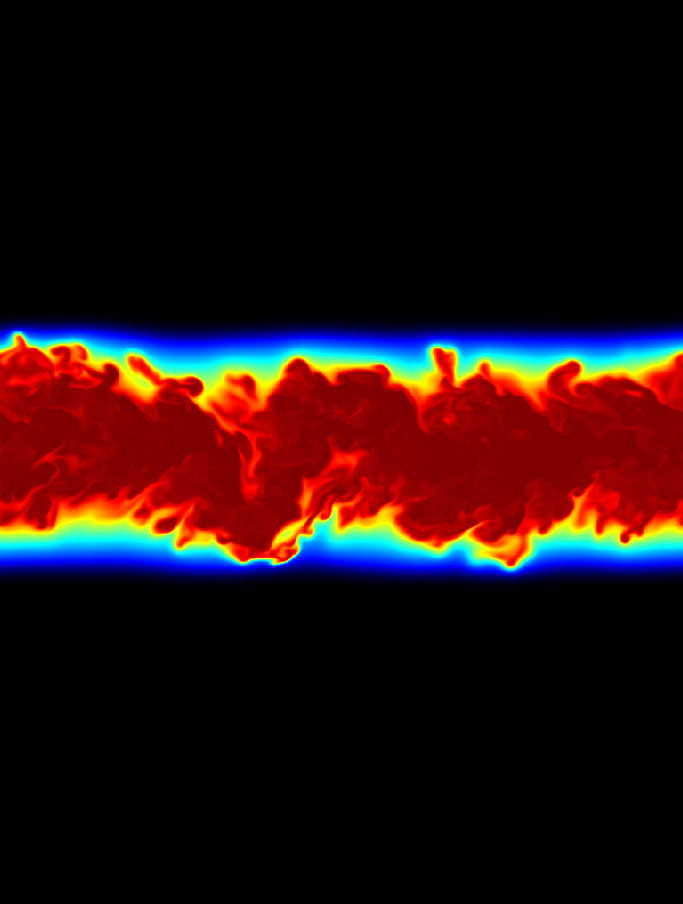}
            \vspace*{-10mm}
            \caption*{\colorbox{white}{$Z_\mathrm{R}$}}  
            \label{dnt:sfig:over4e:1R}
        \end{subfigure}
            \vskip 1mm

        \begin{subfigure}[b]{22mm}
            \centering
            \includegraphics[width=\textwidth]{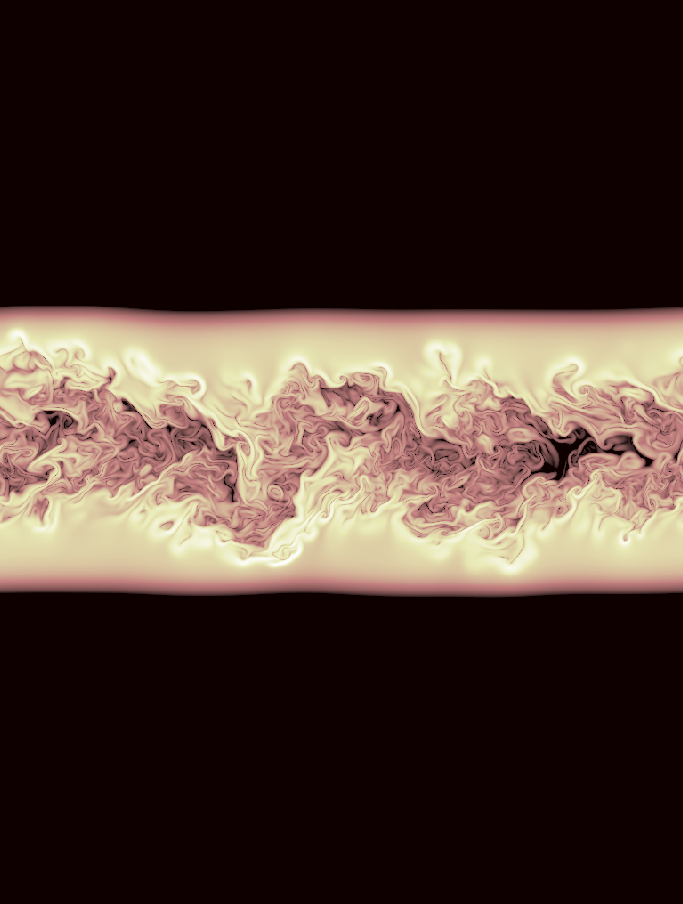}
            \vspace*{-9.5mm}
            \caption*{\colorbox{white}{$\chi_\mathrm{H}$}}  
            \label{dnt:sfig:over4e:2H}
        \end{subfigure}
        \hspace{0mm}
        \begin{subfigure}[b]{22mm}
            \centering
            \includegraphics[width=\textwidth]{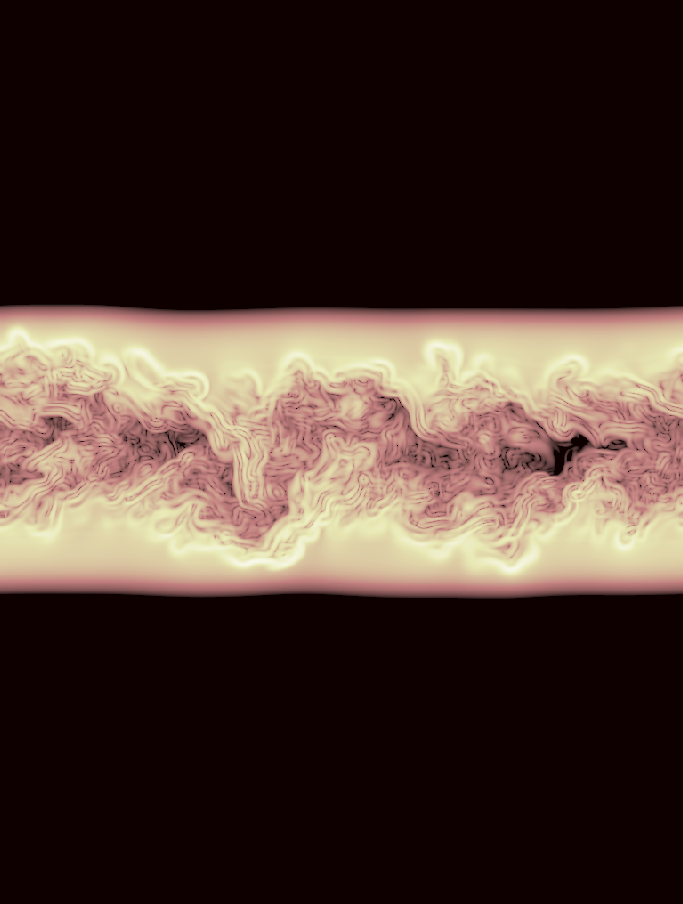}
            \vspace*{-9.5mm}
            \caption*{\colorbox{white}{$\chi_\mathrm{F}$}}  
            \label{dnt:sfig:over4e:2F}
        \end{subfigure}
        \hspace{0mm}
        \begin{subfigure}[b]{22mm}
            \centering
            \includegraphics[width=\textwidth]{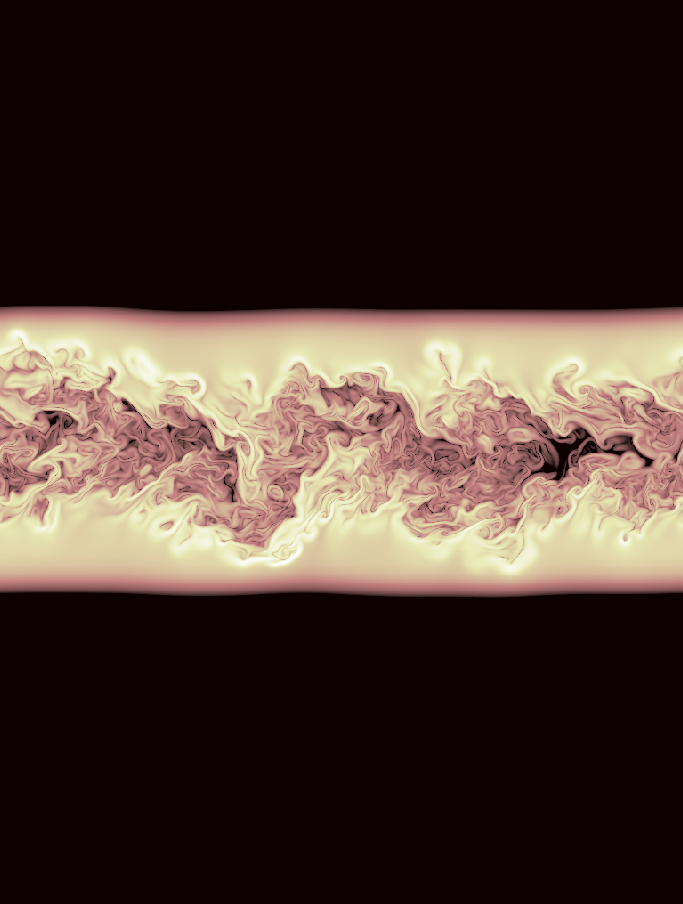}
            \vspace*{-9.5mm}
            \caption*{\colorbox{white}{$\chi_\mathrm{R}$}}  
            \label{dnt:sfig:over4e:2R}
        \end{subfigure}
            \vskip 1mm

        \begin{subfigure}[b]{22mm}
            \centering
            \includegraphics[width=\textwidth]{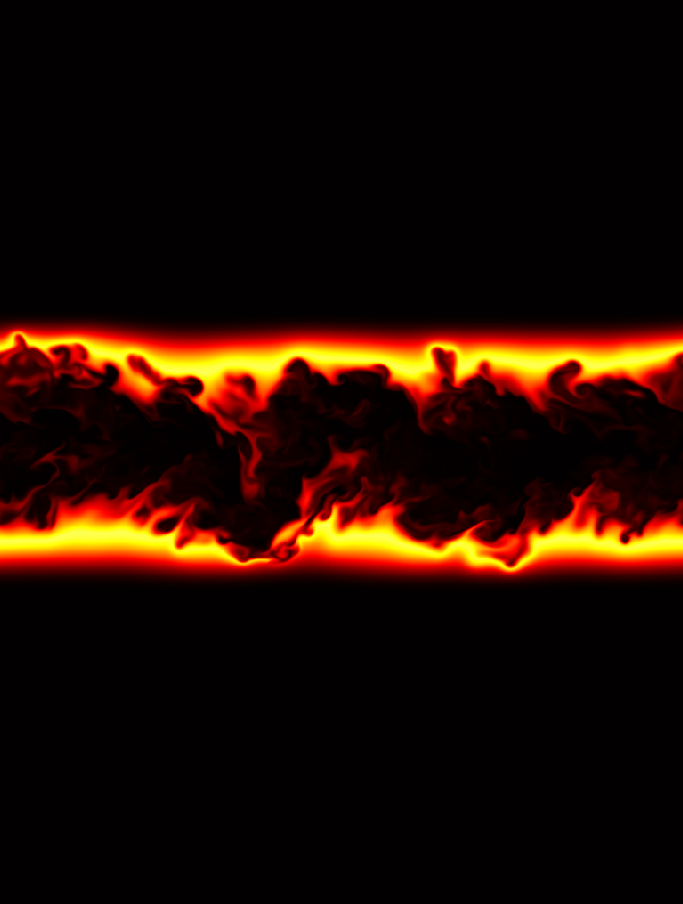}
            \vspace*{-10mm}
            \caption*{\colorbox{white}{$T_\mathrm{H}$}}  
            \label{dnt:sfig:over4e:3H}
        \end{subfigure}
        \hspace{0mm}
        \begin{subfigure}[b]{22mm}
            \centering
            \includegraphics[width=\textwidth]{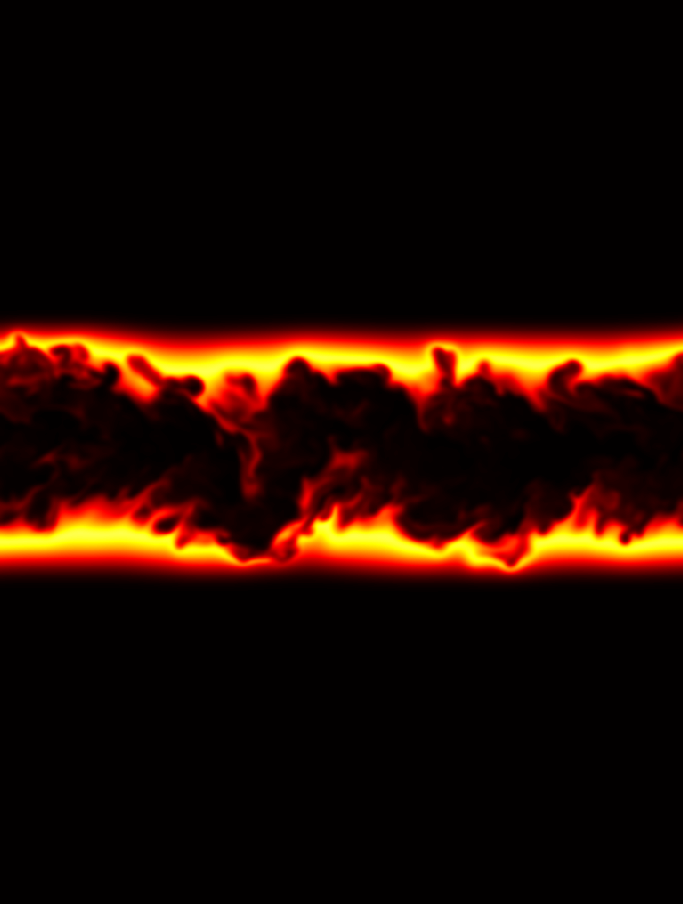}
            \vspace*{-10mm}
            \caption*{\colorbox{white}{$T_\mathrm{F}$}}  
            \label{dnt:sfig:over4e:3F}
        \end{subfigure}
        \hspace{0mm}
        \begin{subfigure}[b]{22mm}
            \centering
            \includegraphics[width=\textwidth]{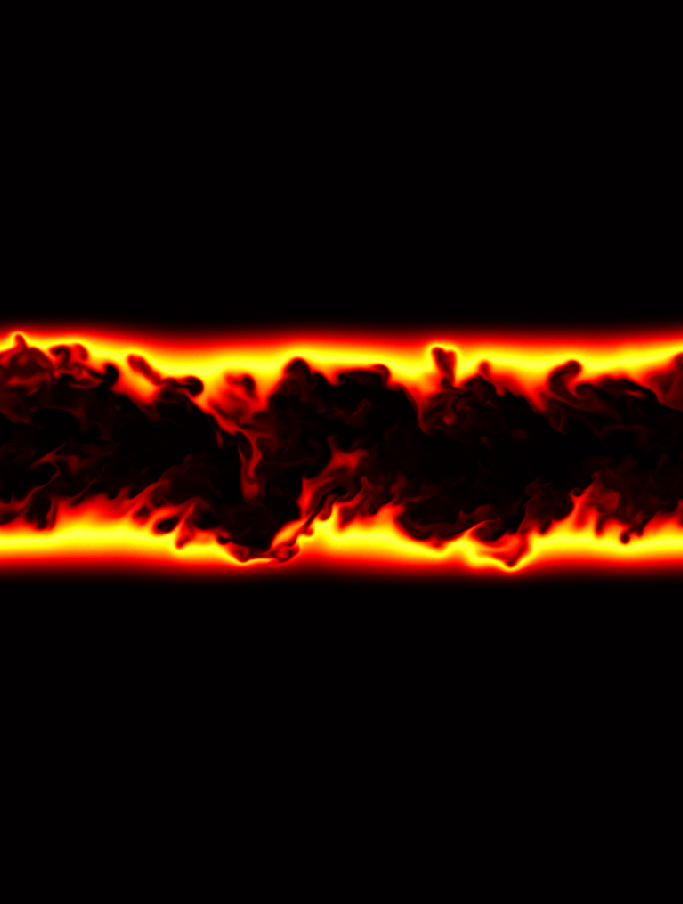}
            \vspace*{-10mm}
            \caption*{\colorbox{white}{$T_\mathrm{R}$}}  
            \label{dnt:sfig:over4e:3R}
        \end{subfigure}
    \caption{LRe early}
    \label{dnt:sfig:over4e}
    \end{subfigure}
    \vskip 3mm

    \begin{subfigure}[b]{\columnwidth}
    \centering
        \begin{subfigure}[b]{22mm}
            \centering
            \includegraphics[width=\textwidth]{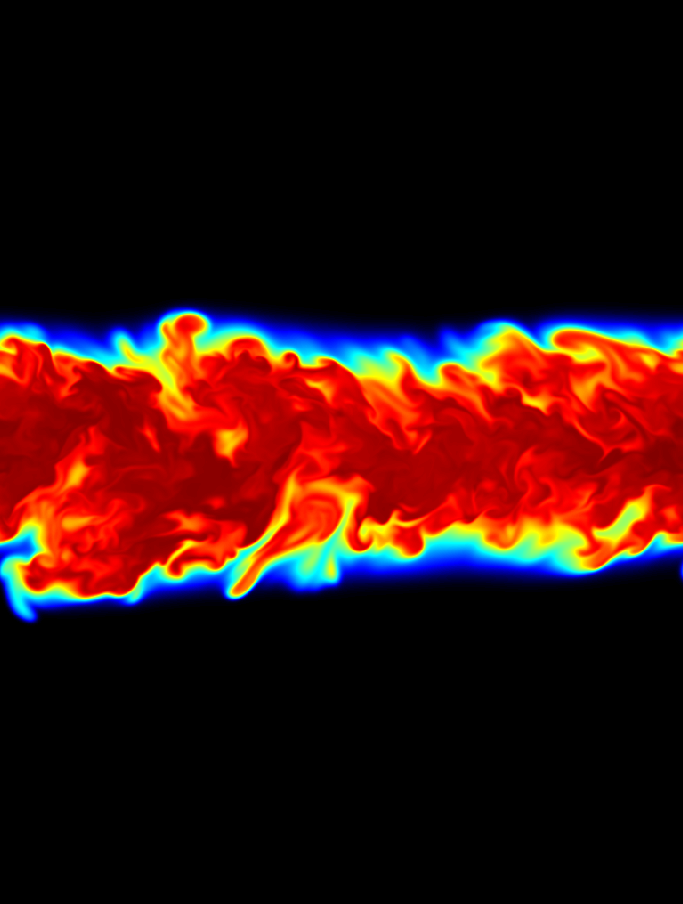}
            \vspace*{-10mm}
            \caption*{\colorbox{white}{$Z_\mathrm{H}$}}  
            \label{dnt:sfig:over4l:1H}
        \end{subfigure}
        \hspace{0mm}
        \begin{subfigure}[b]{22mm}
            \centering
            \includegraphics[width=\textwidth]{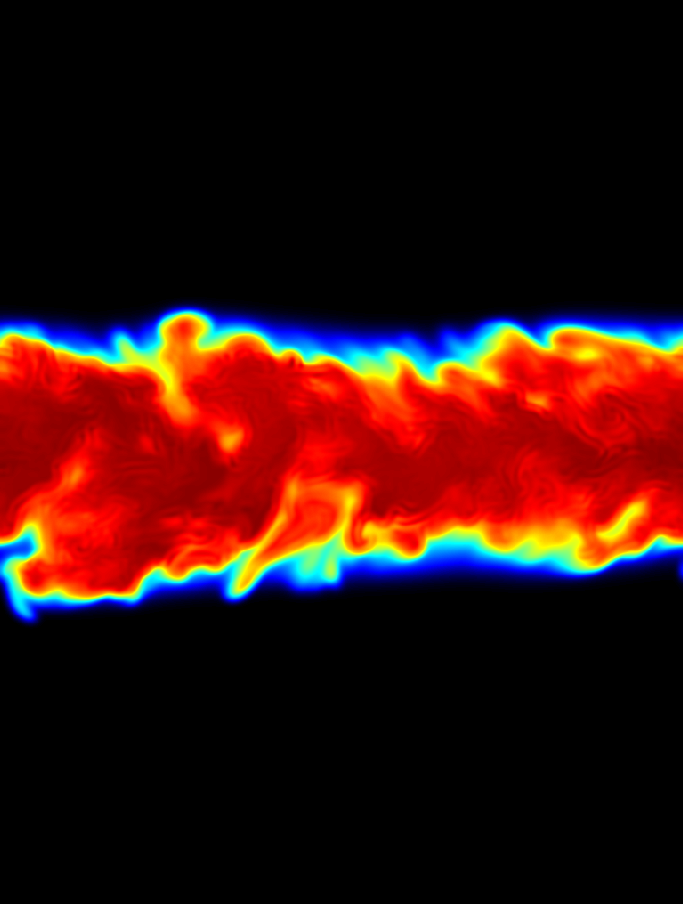}
            \vspace*{-10mm}
            \caption*{\colorbox{white}{$Z_\mathrm{F}$}}  
            \label{dnt:sfig:over4l:1F}
        \end{subfigure}
        \hspace{0mm}
        \begin{subfigure}[b]{22mm}
            \centering
            \includegraphics[width=\textwidth]{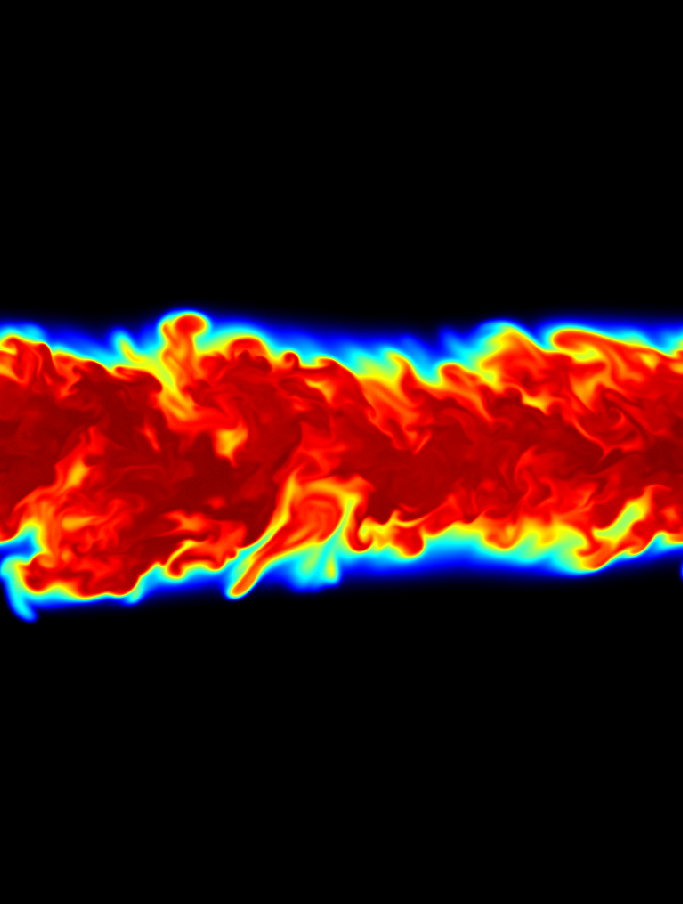}
            \vspace*{-10mm}
            \caption*{\colorbox{white}{$Z_\mathrm{R}$}}  
            \label{dnt:sfig:over4l:1R}
        \end{subfigure}
        \vskip 1mm

        \begin{subfigure}[b]{22mm}
            \centering
            \includegraphics[width=\textwidth]{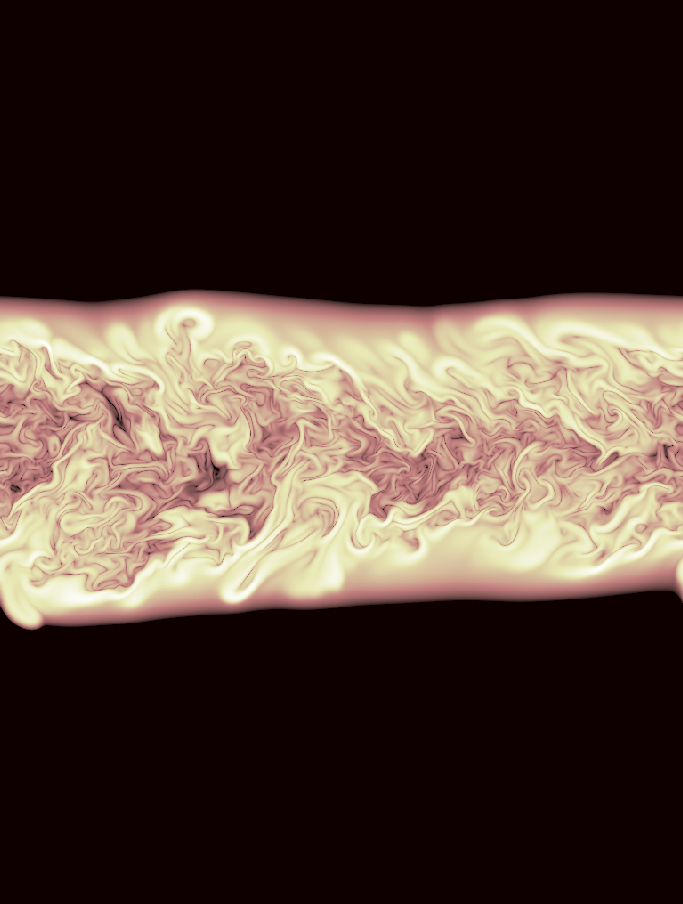}
            \vspace*{-9.5mm}
            \caption*{\colorbox{white}{$\chi_\mathrm{H}$}}  
            \label{dnt:sfig:over4l:2H}
        \end{subfigure}
        \hspace{0mm}
        \begin{subfigure}[b]{22mm}
            \centering
            \includegraphics[width=\textwidth]{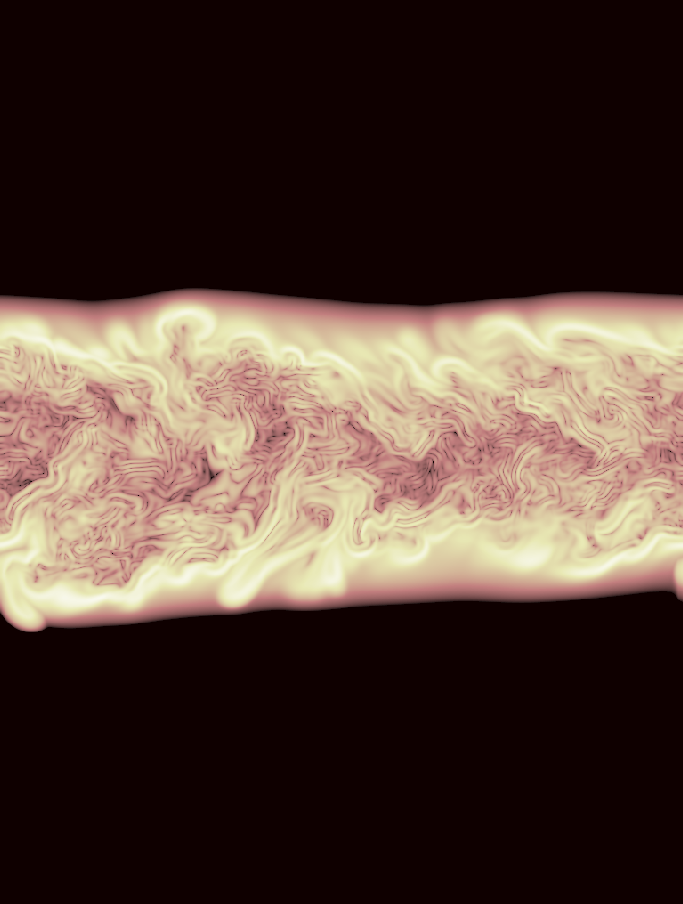}
            \vspace*{-9.5mm}
            \caption*{\colorbox{white}{$\chi_\mathrm{F}$}}  
            \label{dnt:sfig:over4l:2F}
        \end{subfigure}
        \hspace{0mm}
        \begin{subfigure}[b]{22mm}
            \centering
            \includegraphics[width=\textwidth]{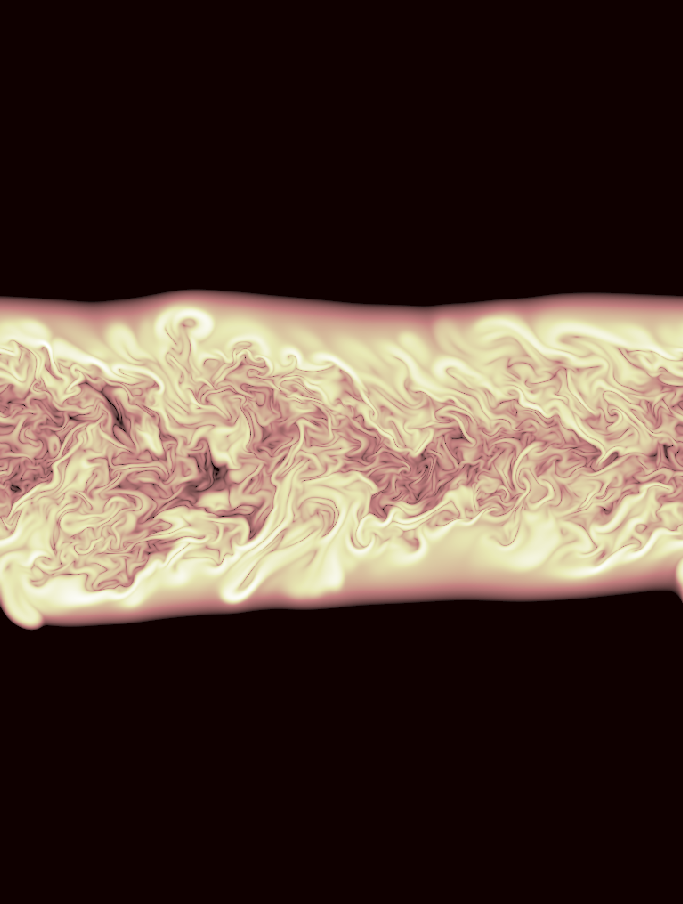}
            \vspace*{-9.5mm}
            \caption*{\colorbox{white}{$\chi_\mathrm{R}$}}  
            \label{dnt:sfig:over4l:2R}
        \end{subfigure}
        \vskip 1mm

        \begin{subfigure}[b]{22mm}
            \centering
            \includegraphics[width=\textwidth]{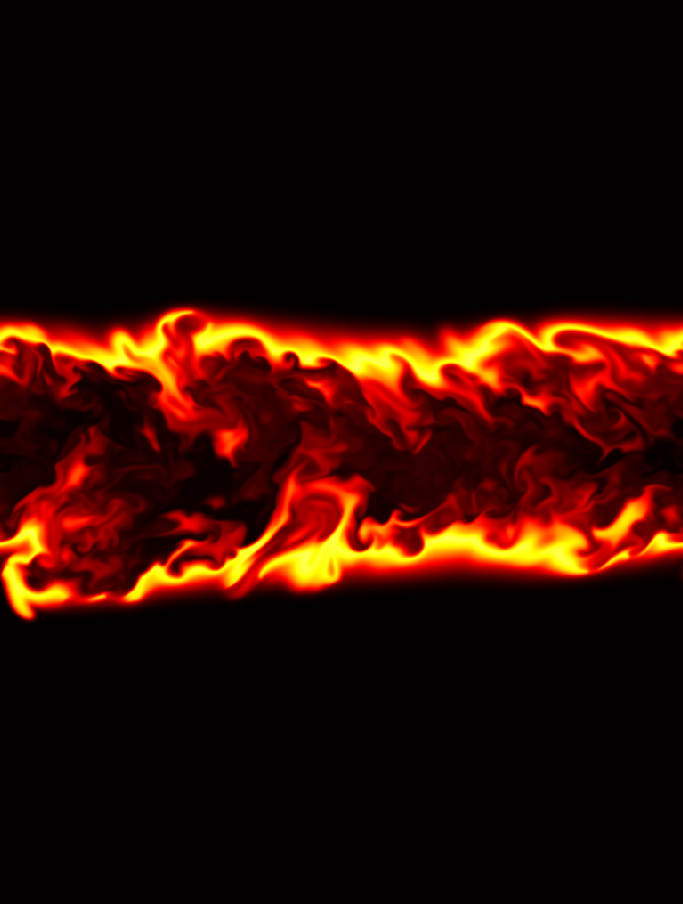}
            \vspace*{-10mm}
            \caption*{\colorbox{white}{$T_\mathrm{H}$}}  
            \label{dnt:sfig:over4l:3H}
        \end{subfigure}
        \hspace{0mm}
        \begin{subfigure}[b]{22mm}
            \centering
            \includegraphics[width=\textwidth]{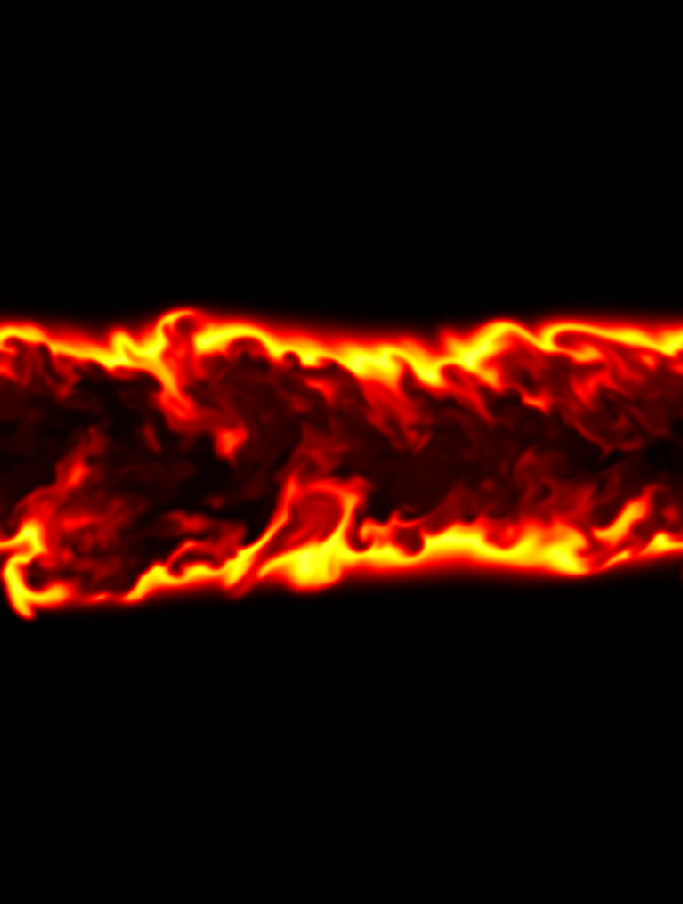}
            \vspace*{-10mm}
            \caption*{\colorbox{white}{$T_\mathrm{F}$}}  
            \label{dnt:sfig:over4l:3F}
        \end{subfigure}
        \hspace{0mm}
        \begin{subfigure}[b]{22mm}
            \centering
            \includegraphics[width=\textwidth]{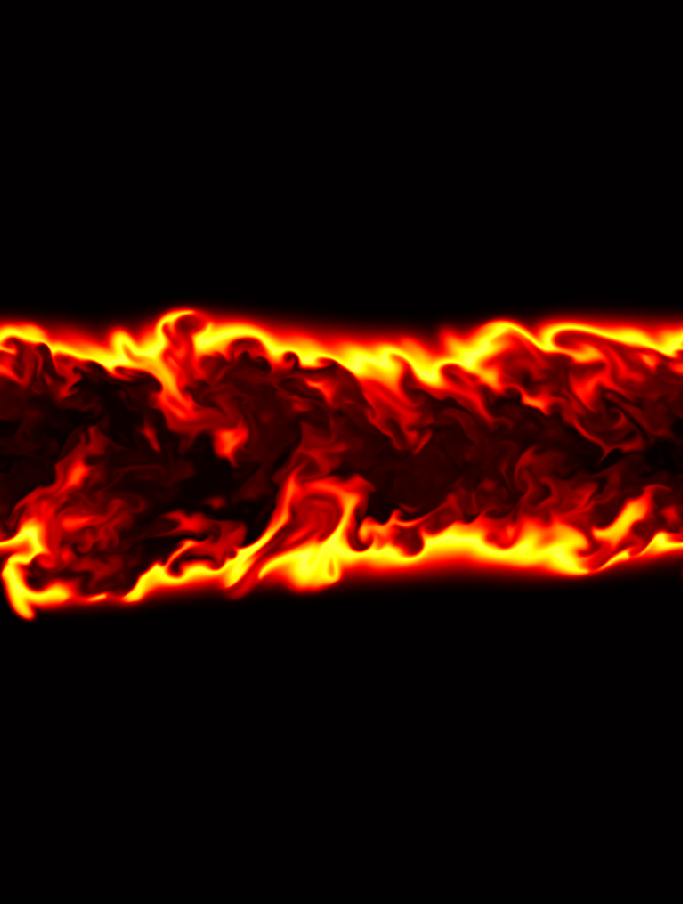}
            \vspace*{-10mm}
            \caption*{\colorbox{white}{$T_\mathrm{R}$}}  
            \label{dnt:sfig:over4l:3R}
        \end{subfigure}
    \caption{LRe late}
    \label{dnt:sfig:over4l}
    \end{subfigure}
       \caption{Visualization of 2-D slices of fully resolved data ("H"), filtered data ("F"), and reconstructed data ("R") of the mixture fraction $Z$, the scalar dissipation rate $\chi$, and the temperature $T$ for LRe at an early and a late time step.}
        \label{dnt:fig:over4}
    \end{figure}

\begin{figure}[!htb]
    \picsize
    \centering
    \begin{subfigure}[b]{\columnwidth}
    \centering
        \begin{subfigure}[b]{22mm}
            \centering
            \includegraphics[width=\textwidth]{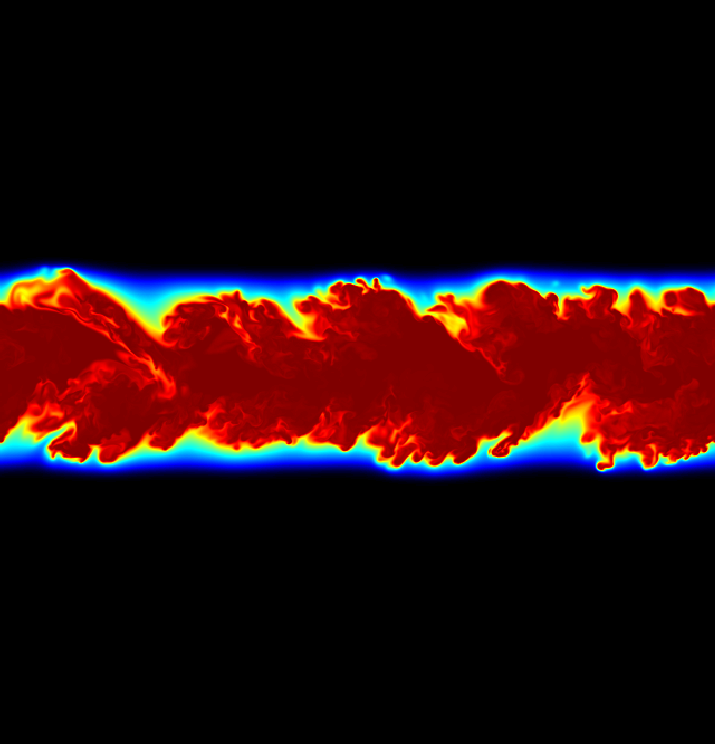}
            \vspace*{-10mm}
            \caption*{\colorbox{white}{$Z_\mathrm{H}$}}  
            \label{dnt:sfig:over1e:1H}
        \end{subfigure}
        \hspace{0mm}
        \begin{subfigure}[b]{22mm}
            \centering
            \includegraphics[width=\textwidth]{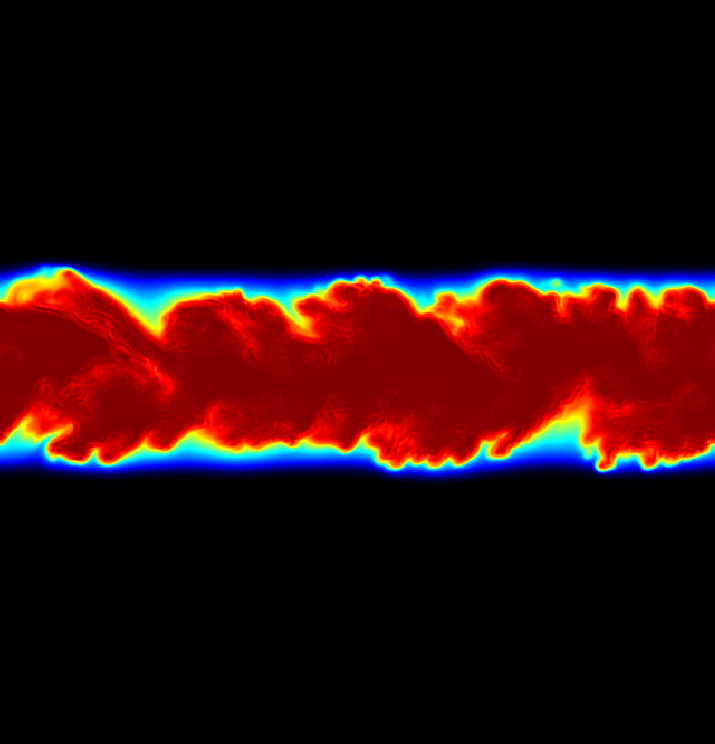}
            \vspace*{-10mm}
            \caption*{\colorbox{white}{$Z_\mathrm{F}$}}  
            \label{dnt:sfig:over1e:1F}
        \end{subfigure}
        \hspace{0mm}
        \begin{subfigure}[b]{22mm}
            \centering
            \includegraphics[width=\textwidth]{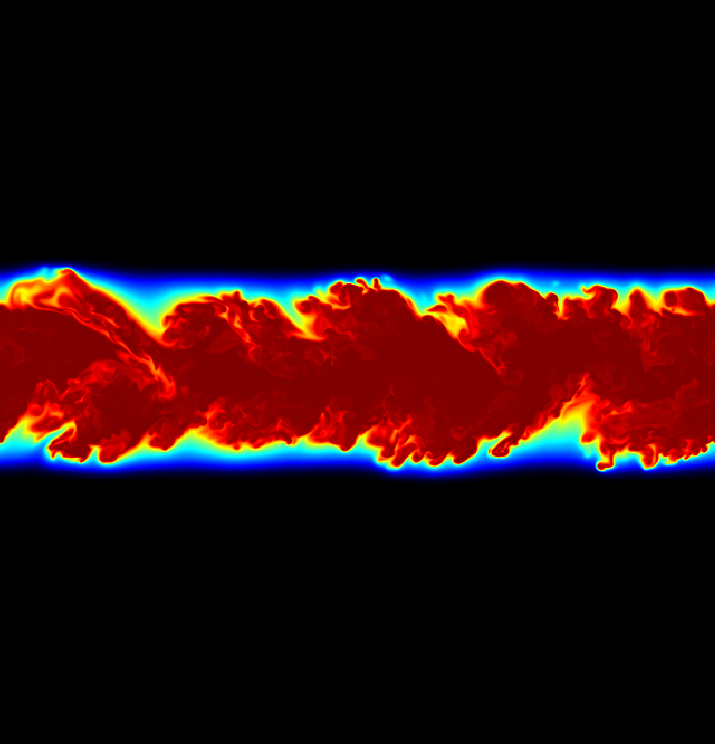}
            \vspace*{-10mm}
            \caption*{\colorbox{white}{$Z_\mathrm{R}$}}  
            \label{dnt:sfig:over1e:1R}
        \end{subfigure}
        \vskip 1mm

        \begin{subfigure}[b]{22mm}
            \centering
            \includegraphics[width=\textwidth]{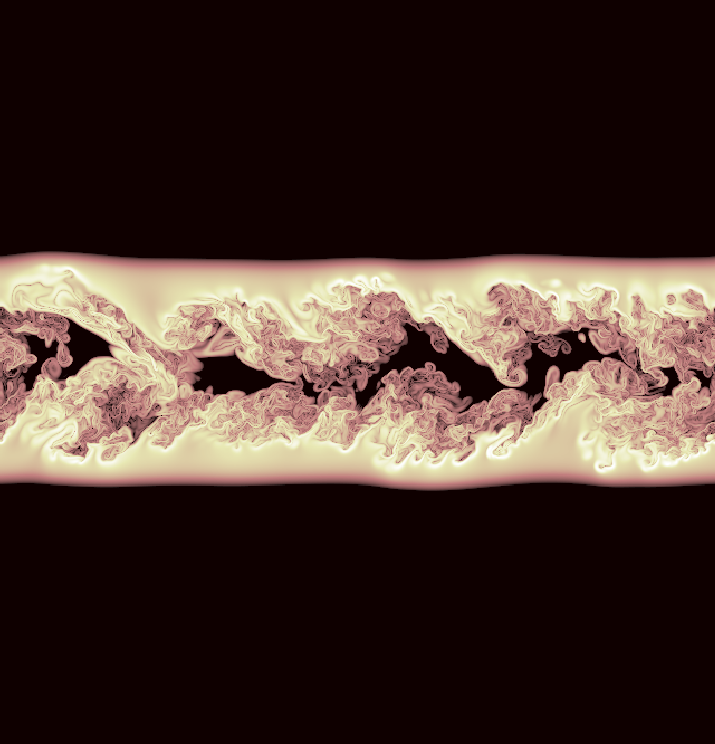}
            \vspace*{-9.5mm}
            \caption*{\colorbox{white}{$\chi_\mathrm{H}$}}  
            \label{dnt:sfig:over1e:2H}
        \end{subfigure}
        \hspace{0mm}
        \begin{subfigure}[b]{22mm}
            \centering
            \includegraphics[width=\textwidth]{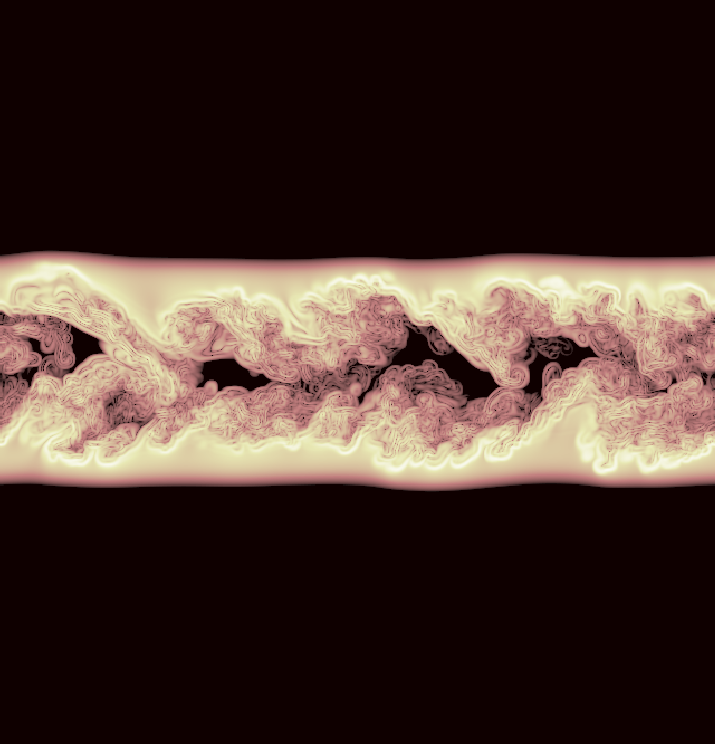}
            \vspace*{-9.5mm}
            \caption*{\colorbox{white}{$\chi_\mathrm{F}$}}  
            \label{dnt:sfig:over1e:2F}
        \end{subfigure}
        \hspace{0mm}
        \begin{subfigure}[b]{22mm}
            \centering
            \includegraphics[width=\textwidth]{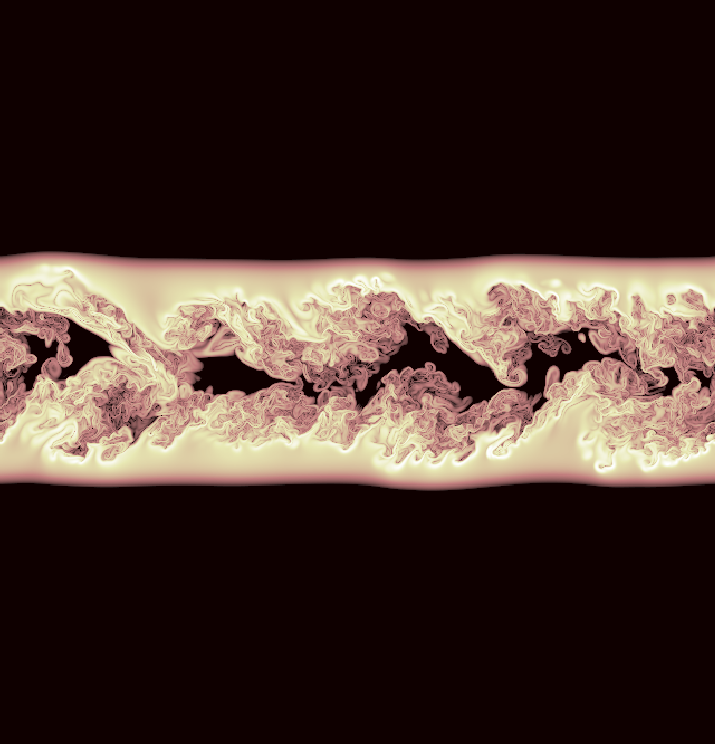}
            \vspace*{-9.5mm}
            \caption*{\colorbox{white}{$\chi_\mathrm{R}$}}  
            \label{dnt:sfig:over1e:2R}
        \end{subfigure}
        \vskip 1mm

        \begin{subfigure}[b]{22mm}
            \centering
            \includegraphics[width=\textwidth]{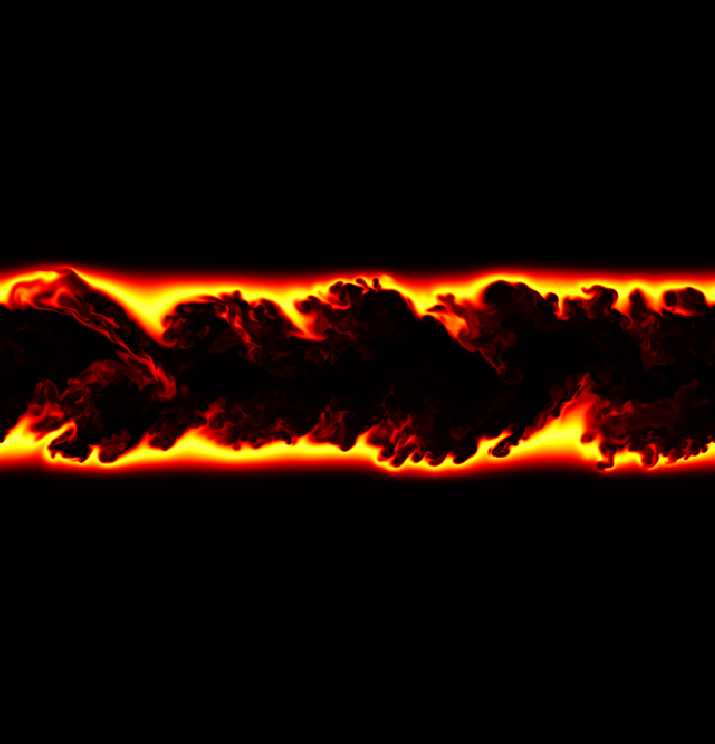}
            \vspace*{-10mm}
            \caption*{\colorbox{white}{$T_\mathrm{H}$}}  
            \label{dnt:sfig:over1e:3H}
        \end{subfigure}
        \hspace{0mm}
        \begin{subfigure}[b]{22mm}
            \centering
            \includegraphics[width=\textwidth]{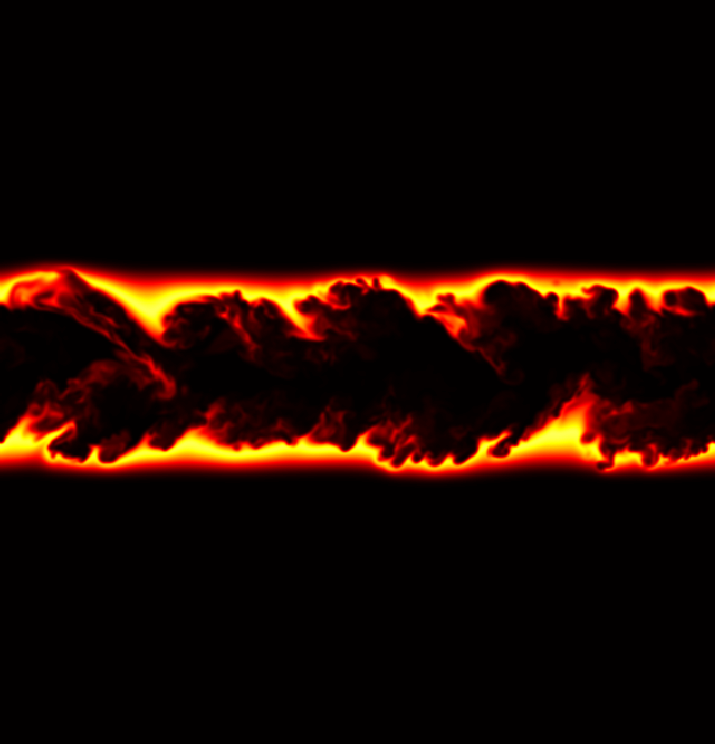}
            \vspace*{-10mm}
            \caption*{\colorbox{white}{$T_\mathrm{F}$}}  
            \label{dnt:sfig:over1e:3F}
        \end{subfigure}
        \hspace{0mm}
        \begin{subfigure}[b]{22mm}
            \centering
            \includegraphics[width=\textwidth]{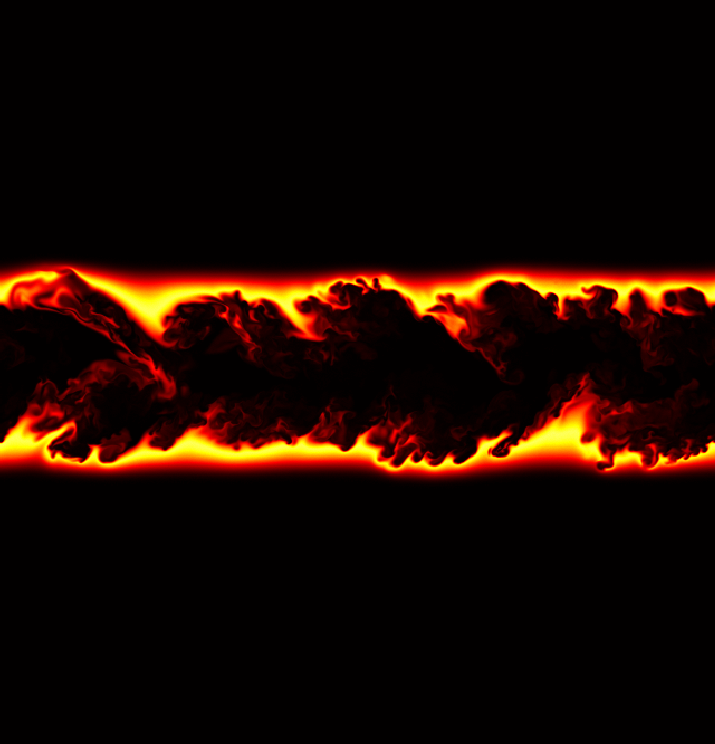}
            \vspace*{-10mm}
            \caption*{\colorbox{white}{$T_\mathrm{R}$}}  
            \label{dnt:sfig:over1e:3R}
        \end{subfigure}
    \caption{HRe early}
    \label{dnt:sfig:over1e}
    \end{subfigure}
    \vskip 3mm

    \begin{subfigure}[b]{\columnwidth}
    \centering
        \begin{subfigure}[b]{22mm}
            \centering
            \includegraphics[width=\textwidth]{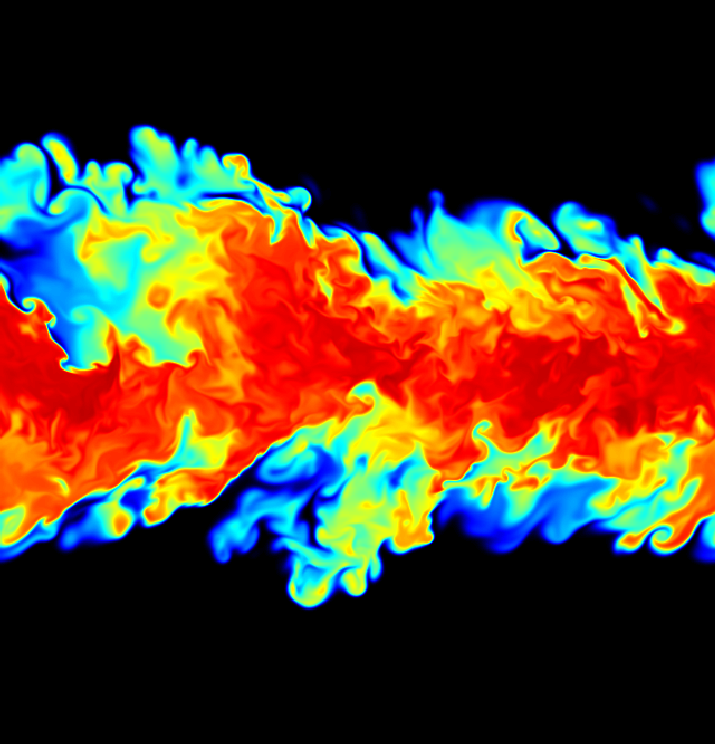}
            \vspace*{-10mm}
            \caption*{\colorbox{white}{$Z_\mathrm{H}$}}  
            \label{dnt:sfig:over1l:1H}
        \end{subfigure}
        \hspace{0mm}
        \begin{subfigure}[b]{22mm}
            \centering
            \includegraphics[width=\textwidth]{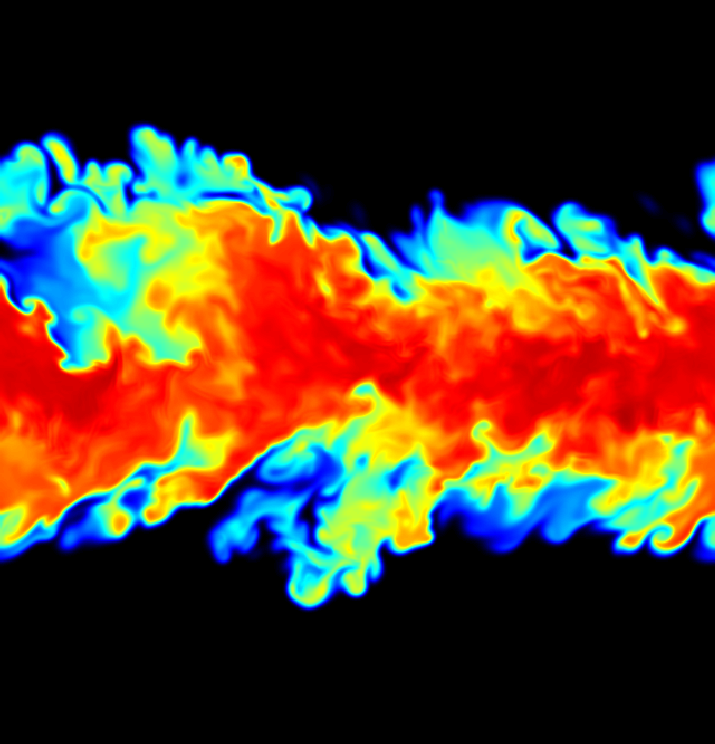}
            \vspace*{-10mm}
            \caption*{\colorbox{white}{$Z_\mathrm{F}$}}  
            \label{dnt:sfig:over1l:1F}
        \end{subfigure}
        \hspace{0mm}
        \begin{subfigure}[b]{22mm}
            \centering
            \includegraphics[width=\textwidth]{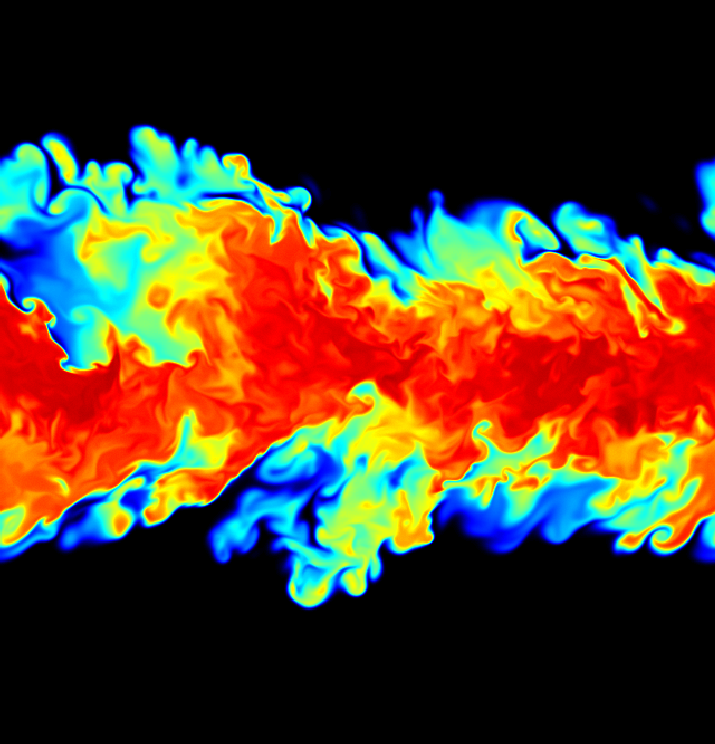}
            \vspace*{-10mm}
            \caption*{\colorbox{white}{$Z_\mathrm{R}$}}  
            \label{dnt:sfig:over1l:1R}
        \end{subfigure}
        \vskip 1mm

        \begin{subfigure}[b]{22mm}
            \centering
            \includegraphics[width=\textwidth]{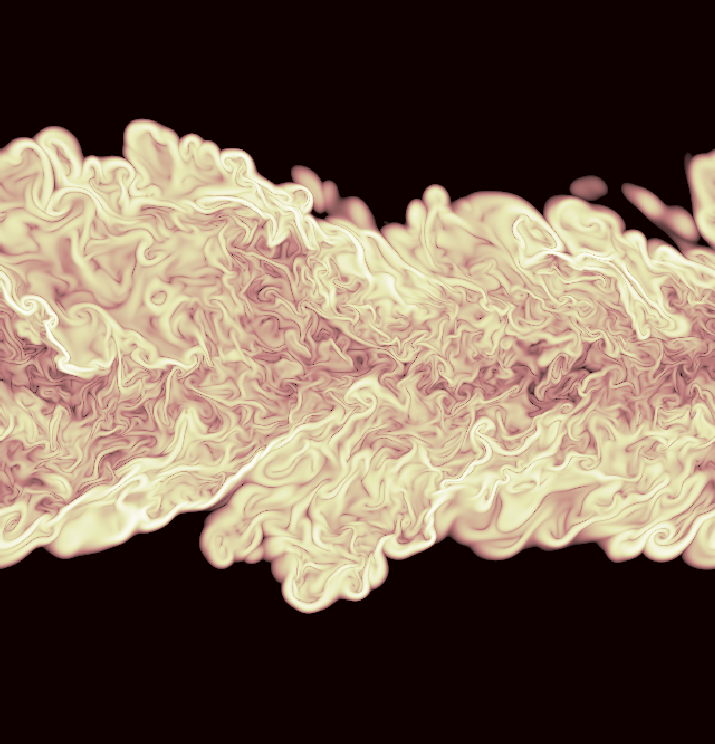}
            \vspace*{-9.5mm}
            \caption*{\colorbox{white}{$\chi_\mathrm{H}$}}  
            \label{dnt:sfig:over1l:2H}
        \end{subfigure}
        \hspace{0mm}
        \begin{subfigure}[b]{22mm}
            \centering
            \includegraphics[width=\textwidth]{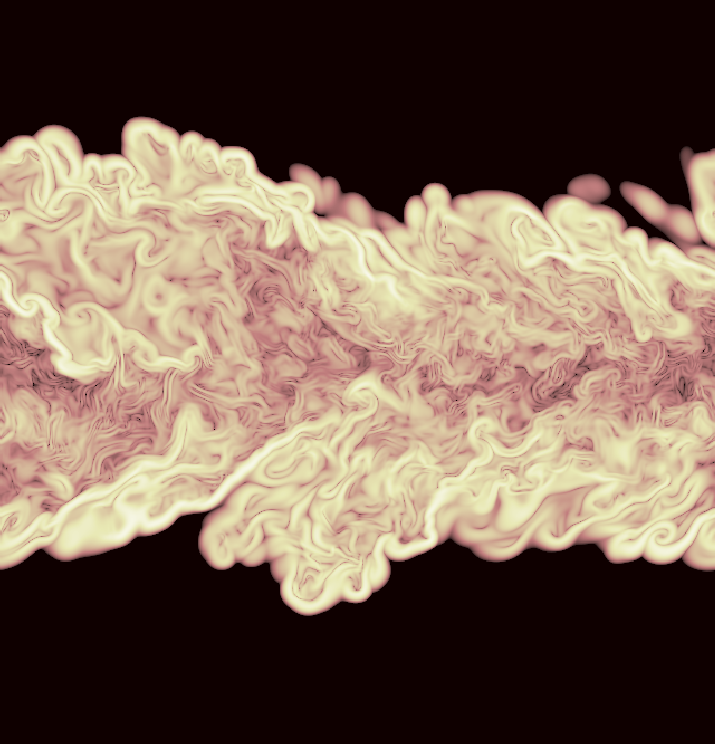}
            \vspace*{-9.5mm}
            \caption*{\colorbox{white}{$\chi_\mathrm{F}$}}  
            \label{dnt:sfig:over1l:2F}
        \end{subfigure}
        \hspace{0mm}
        \begin{subfigure}[b]{22mm}
            \centering
            \includegraphics[width=\textwidth]{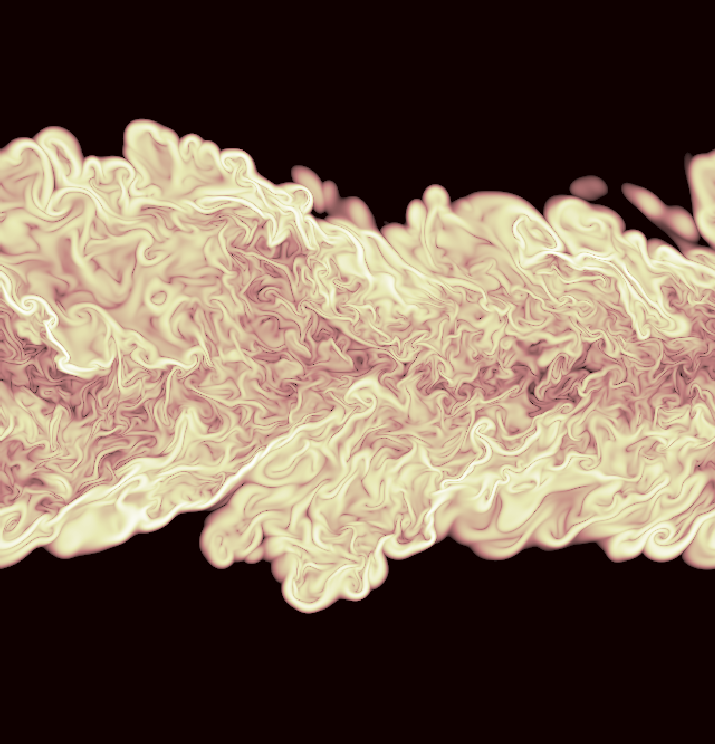}
            \vspace*{-9.5mm}
            \caption*{\colorbox{white}{$\chi_\mathrm{R}$}}  
            \label{dnt:sfig:over1l:2R}
        \end{subfigure}
        \vskip 1mm

        \begin{subfigure}[b]{22mm}
            \centering
            \includegraphics[width=\textwidth]{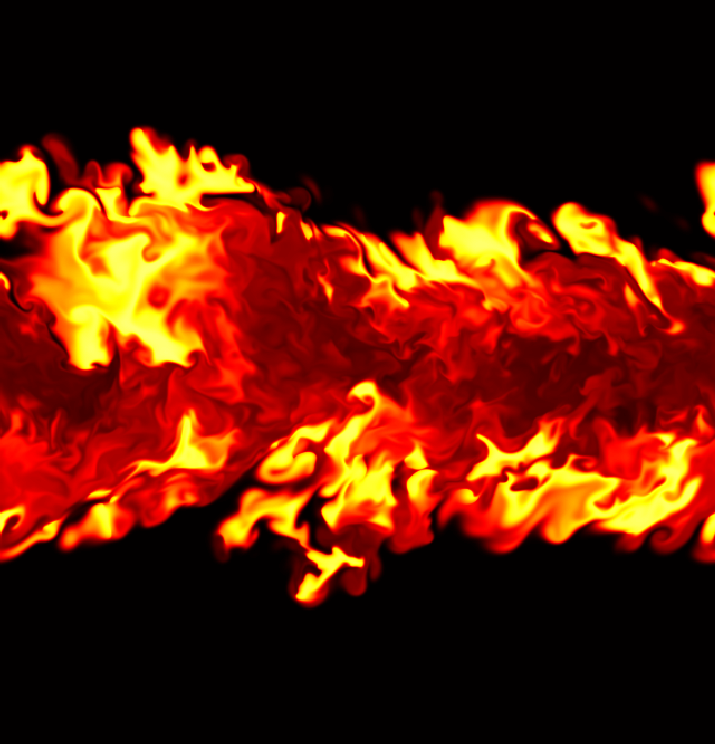}
            \vspace*{-10mm}
            \caption*{\colorbox{white}{$T_\mathrm{H}$}}  
            \label{dnt:sfig:over1l:3H}
        \end{subfigure}
        \hspace{0mm}
        \begin{subfigure}[b]{22mm}
            \centering
            \includegraphics[width=\textwidth]{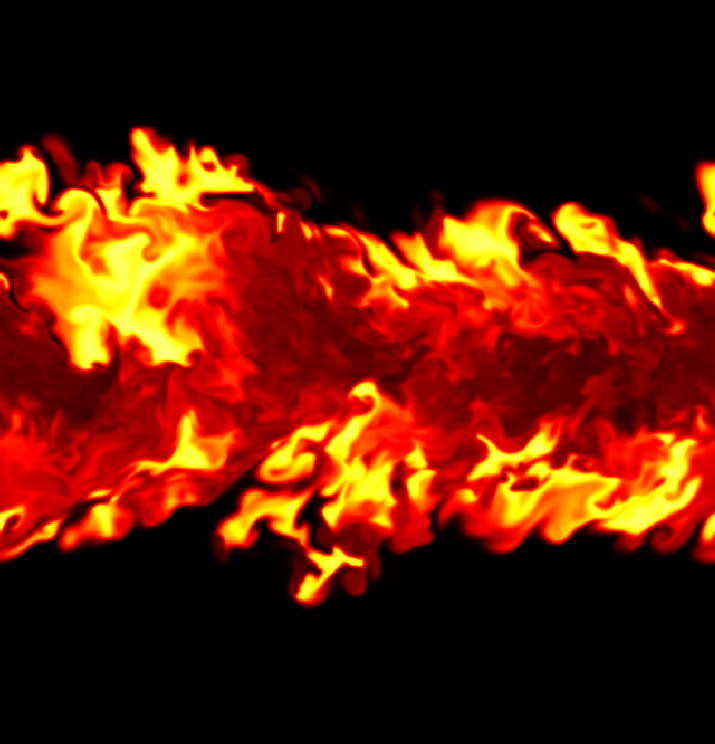}
            \vspace*{-10mm}
            \caption*{\colorbox{white}{$T_\mathrm{F}$}}  
            \label{dnt:sfig:over1l:3F}
        \end{subfigure}
        \hspace{0mm}
        \begin{subfigure}[b]{22mm}
            \centering
            \includegraphics[width=\textwidth]{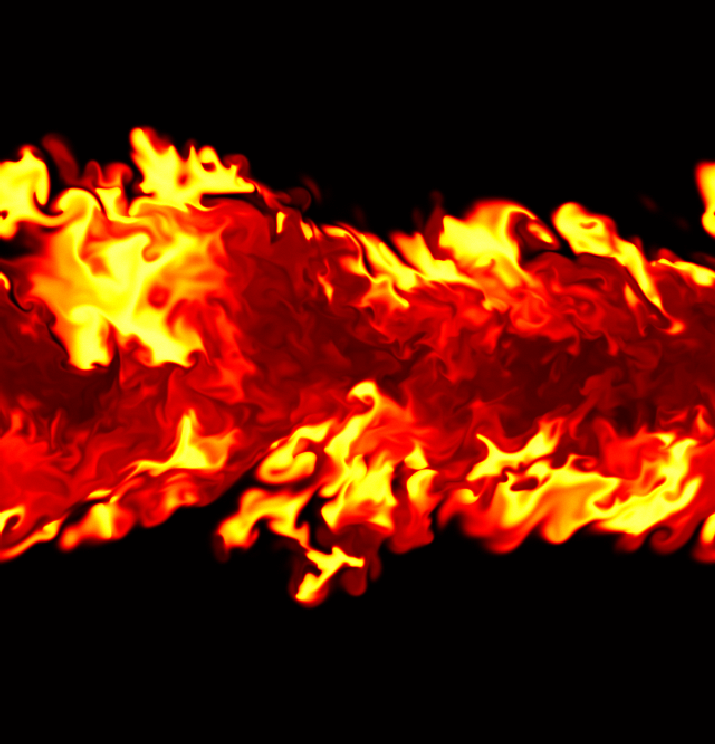}
            \vspace*{-10mm}
            \caption*{\colorbox{white}{$T_\mathrm{R}$}}  
            \label{dnt:sfig:over1l:3R}
        \end{subfigure}
    \caption{HRe late}
    \label{dnt:sfig:over1l}
    \end{subfigure}
       \caption{Visualization of 2-D slices of fully resolved data ("H"), filtered data ("F"), and reconstructed data ("R") of the mixture fraction $Z$, the scalar dissipation rate $\chi$, and the temperature $T$ for HRe at an early and a late time step.}
        \label{dnt:fig:over1}
    \end{figure}

To get a first qualitative impression of the PG prediction and the evolution of the cases themselves, as well as the differences between the cases, \reff{dnt:fig:over4} and \reff{dnt:fig:over1} show two time snapshots for each case. 2-D cuts of the mixture fraction $Z$, the scalar dissipation rate $\chi$, and the temperature $T$ are shown in three versions: the fully resolved data from DNSs, the filtered data, and the reconstructed data computed with PGs using the filtered data as input. It is notable that the first two fields are not directly solved in the simulations/reconstructions. Instead they are computed by means of the mass fraction fields. Both cases show an increased mixing of fuel and oxidizer stream over time. The turbulent area strongly increases in $x_2$ direction over time. As expected, the mixing is much stronger in the case with high Reynolds numbers. The filtered data represent data filtered with a medium sized filter stencil width. Even though the filtering does not change the main character of the flow, it is obvious that small scale structures vanish by smoothing. The visual agreement between fully resolved data and reconstructed data is very good. 

A quantitative assessment of the prediction quality is discussed in the next sections by means of PDFs. All PDFs show a very strong peak at zero, representing, e.\,g., inert or laminar regions. This peak is not shown in the plots by choosing the ordinate accordingly.

\subsubsection{Mixture fraction}
The mixture fraction is one of the main indicators in non-premixed combustion. As fuel and oxidizer are originally separated, mixing must take place before chemistry can happen. The mixture fraction was neither directly solved in the DNS nor during the reconstruction. Instead, it was computed by means of the mass fraction fields, which were solved and reconstructed, respectively.

\reff{dnt:fig:pdfz} shows the PDFs for the mixture fractions for LRe and HRe at two time steps each indicated by the used symbols: triangles mark the earlier time step and boxes mark the later time step. Filtering results with three different filter stencil widths are shown with $\Delta_1 < \Delta_2 < \Delta_3$.  The qualitative temporal evolution is similar for both cases. The peak of the PDFs shifts from $Z=1$ to smaller values as mixing occurs. This process is further for the HRe case, which is expected by the higher Reynolds number. However, also note that the cases do not exactly represent the same time. The effect of the filter is visible but not too strong. The larger the filter width, the stronger are the differences, as especially visible for the later time step of LRe. The reconstruction result is very good.
\begin{figure}[!tb]
\picsize
	\centering
        \begin{subfigure}[b]{\columnwidth}
            \centering
            \picbox{\input{figures/dnt_pdf_z_tex_4500.tex}}
            \caption{LRe}  
            \label{dnt:sfig:pdfz:1}
        \end{subfigure}
        
        \begin{subfigure}[b]{\columnwidth}
            \centering
            \picbox{\input{figures/dnt_pdf_z_tex_10000.tex}}
            \caption{HRe}  
            \label{dnt:sfig:pdfz:2}
        \end{subfigure}
	\caption{PDFs of the mixture fraction $Z$ for LRe and HRe at an earlier time step (triangles) and a later time step (boxes). PDFs evaulated on the DNS data, filtered data (three different kernel widths), and reconstructed data are shown.}
	\label{dnt:fig:pdfz}
\end{figure}

\subsubsection{Scalar dissipation rate}
\begin{figure}[!tb]
\picsize
	\centering
        \begin{subfigure}[b]{\columnwidth}
            \centering
            \picbox{\input{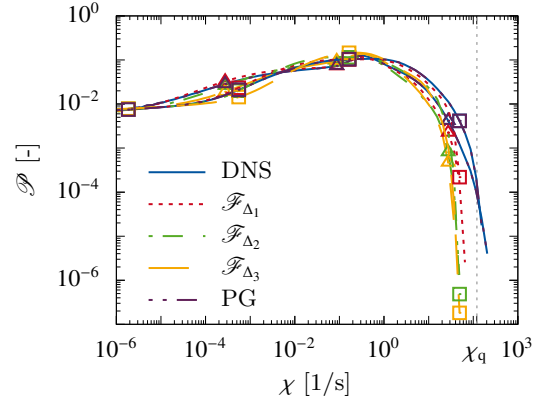}}
            \caption{LRe}  
            \label{dnt:sfig:pdfchi:1}
        \end{subfigure}
        
        \begin{subfigure}[b]{\columnwidth}
            \centering
            \picbox{\input{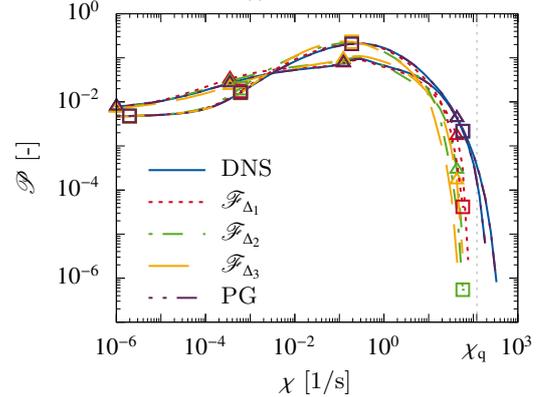}}
            \caption{HRe}  
            \label{dnt:sfig:pdfchi:2}
        \end{subfigure}
	\caption{PDFs of the scalar dissipation rate $\chi$ for LRe and HRe at an earlier time step (triangles) and a later time step (boxes). PDFs evaulated on the DNS data, filtered data (three different kernel widths), and reconstructed data are shown.}
	\label{dnt:fig:pdfchi}
\end{figure}

The scalar dissipation describes the local rate of mixing. Physically, it is very important as quenching occurs for too large scalar dissipation rates, as the mixing is too fast. For the conditions considered here, quenching happens for scalar dissipation rates larger than $\chi_\mathrm{q}=\SI{120}{\per\second}$. The scalar dissipation rate is numerically very challenging: First, it shows multi-scale behavior, and thus, multiple scales need to be predicted correctly. Secondly, the scalar dissipation rate is a derived quantity, and more importantly, is proportional to the squared gradient of the mixture fraction, which itself relies on all mass fractions. The gradient is very difficult to predict, as already small local deviations in the mixture fraction lead to potentially large deviations in the scalar dissipation rate. The PDFs are shown as log-log plot in \reff{dnt:fig:pdfchi} for both cases, LRe and HRe. The filtering removes turbulent fluctuations, which can be clearly seen by the lack of very high scalar dissipation rates in the filtered data. In terms of quenching, this is crucial. The indicated quenching dissipation rate reveals that quenching would not occur in the filtered data but in the DNS, and thus the underlying physics are changed. Again, the reconstruction seems to be good for all cases, however, it should be noted that small deviations are difficult to see due to the log-log representation.

\subsubsection{Species prediction}
\begin{figure}[!tb]
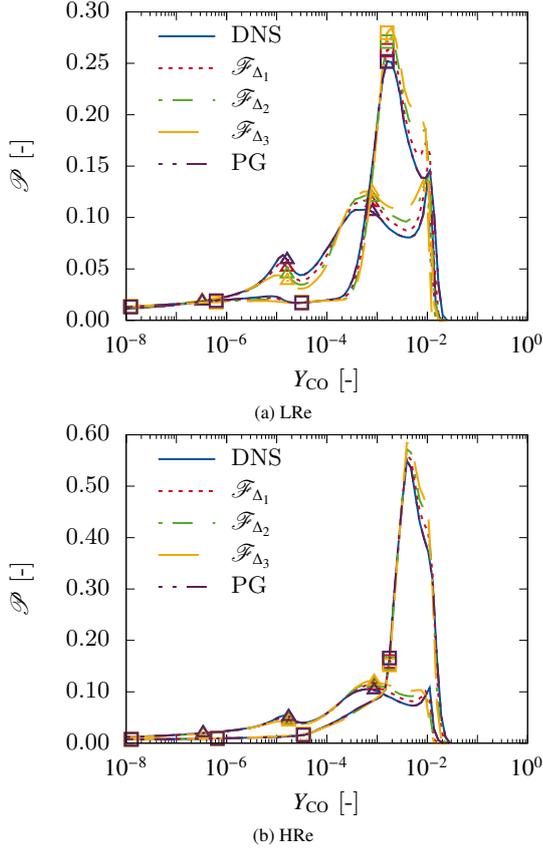

\picsize
	\centering
        \begin{subfigure}[b]{\columnwidth}
            \centering
            \picbox{\input{figures/dnt_pdf_y10_tex_4500.tex}}
            \caption{LRe}  
            \label{dnt:sfig:pdfy10:1}
        \end{subfigure}
        
        \begin{subfigure}[b]{\columnwidth}
            \centering
            \picbox{\input{figures/dnt_pdf_y10_tex_10000.tex}}
            \caption{HRe}  
            \label{dnt:sfig:pdfy10:2}
        \end{subfigure}
	\caption{PDFs of the mass fraction of CO for LRe and HRe at an earlier time step (triangles) and a later time step (boxes). PDFs evaulated on the DNS data, filtered data (three different kernel widths), and reconstructed data are shown.}
	\label{dnt:fig:pdfy10}
\end{figure}

As last a priori test, the PDFs of the mass fractions of two different species are shown in \reff{dnt:fig:pdfy10} and \reff{dnt:fig:pdfy15} for LRe and HRe at two time steps each. While CO is a primary species for the PG reconstruction, $\mathrm{CH}_2\mathrm{O}$ is a secondary species. CO increases towards higher mass fraction values over time, as can be clearly seen in all PDFs. The behavior for $\mathrm{CH}_2\mathrm{O}$ is qualitatively different. While it evolves from a PDF with two smaller peaks at the earlier time step of HRe to a PDF with one distinctive peak at the later time step, this evolution cannot be seen for LRe. However, the two peaks of the PDF in the shown ordinate increase significantly. As before, the agreement between DNS data and reconstructed data is very good for all cases. 
\begin{figure}[!tb]
\picsize
	\centering
        \begin{subfigure}[b]{\columnwidth}
            \centering
            \picbox{\input{figures/dnt_pdf_y15_tex_4500.tex}}
            \caption{LRe}  
            \label{dnt:sfig:pdfy15:1}
        \end{subfigure}
        
        \begin{subfigure}[b]{\columnwidth}
            \centering
            \picbox{\input{figures/dnt_pdf_y15_tex_10000.tex}}
            \caption{HRe}  
            \label{dnt:sfig:pdfy15:2}
        \end{subfigure}
	\caption{PDFs of the mass fraction of $\mathrm{CH}_2\mathrm{O}$ for LRe and HRe at an earlier time step (triangles) and a later time step (boxes). PDFs evaulated on the DNS data, filtered data (three different kernel widths), and reconstructed data are shown.}
	\label{dnt:fig:pdfy15}
\end{figure}

\subsection{A posteriori testing}
The a priori results in the previous section demonstrated that PG is able to very accurately reconstruct the DNS data using only the filtered data as input. However, high accuracy in a priori tests is much simpler to achieve than good accuracy in a posteriori tests. Even small errors in a priori, which are hardy visible, can accumulate over time in a posteriori tests or blow-up the simulation. Therefore, it is highly recommended to introduce also a posteriori metrics in the training process. Fully optimizing the accuracy in an a priori sense only often does not lead to good a posteriori results in the long run.

All a posteriori results discussed here used all three filter stencil widths to correct for the non-uniform mesh. The a posteriori simulations were initialized with the DNS results at $t^{*}=5$. I.\,e., the a posteriori tests were initialized with a rather "stable" state and the more fluctuating initial phase was skipped.

\subsubsection{Surface area}
\begin{figure}[!tb]
\picsize
	\centering
    \picbox{\input{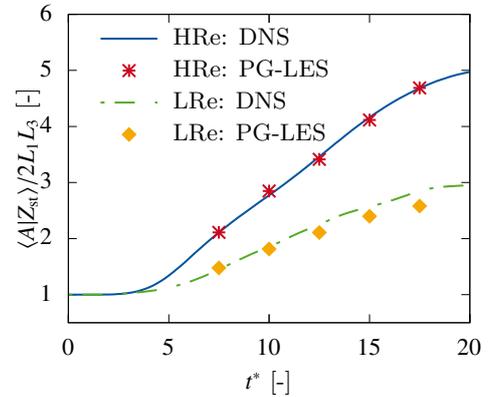}}
	\caption{Evolution of the normalized area of iso-surface with stoichiometric mixture fraction $\langle A \vert Z_\mathrm{st} \rangle/2 L_1 L_3$ over normalized time $t^{*}$ for LRe and HRe from DNS and PG-LES.}
	\label{dnt:fig:ta}
\end{figure}

For both cases, the mixing increases the iso-surface with stoichiometric mixture fraction over time, as turbulence and mixing lead to rugged structures. As this iso-surface is where the oxidation takes place, i.\,e., where the flame sits, it is a very important metric to evaluate the accuracy of the PG-LES. \reff{dnt:fig:ta} shows the temporal evolution of the normalized iso-surface with stoichiometric mixture fraction. The normalization was done with the initial area $2\overline{L}_1\overline{L}_3$. The agreement for HRe is very good. The temporal evolution of LRe is too weak, leading to an underpredicted iso-surface area. It was not possible to identify a reason for the difference, as the level of turbulence was correctly predicted, but it shows how fragile the complex interplay between DNS data and PG-LES is.

\subsubsection{Scalar dissipation rate}
\begin{figure}[!tb]
\picsize
	\centering
    \picbox{\input{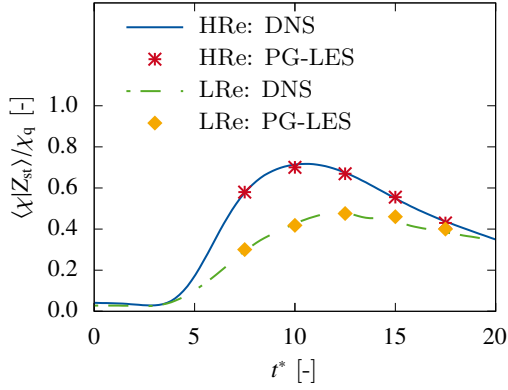}}
	\caption{Evolution of the normalized averaged scalar dissipation rate at locations with stoichiometric mixture fraction $\langle \chi\vert Z_\mathrm{st} \rangle/\chi_\mathrm{q}$ over normalized time $t^{*}$ for LRe and HRe from DNS and PG-LES.}
	\label{dnt:fig:tchi}
\end{figure}

To further evaluate the a posteriori accuracy of the presented method, \reff{dnt:fig:tchi} shows the temporal evolution of the normalized scalar dissipation rate averaged over all locations with stoichiometric mixture fraction for LRe and HRe, which is an important quantity to predict quenching. The prediction accuracy for HRe is still very good, even though not as good as for the area prediction in \reff{dnt:fig:ta}, emphasizing the complex nature of this metric. In contrast, the prediction accuracy is slightly better for LRe compared to the area prediction. However, at later times, the deviation in the area prediction result in a slight overprediction here, as the underlying iso-surface of stoichiometric mixture is not accurately predicted. Overall, the PG-LES prediction quality is very good and much better than what could have been expected with classical LES models. 

%
%

\subsubsection{Effect of primary species}
\begin{figure}[!tb]
\picsize
	\centering
    \picbox{\input{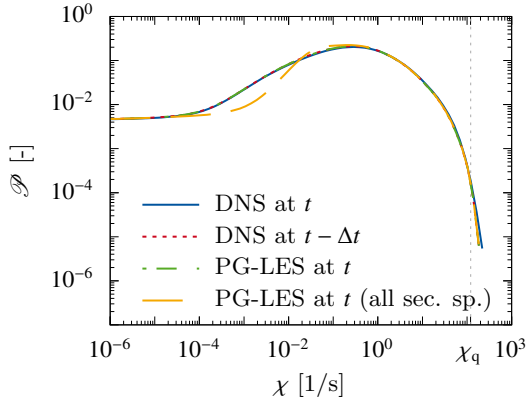}}
	\caption{PDFs of the scalar dissipation rate evaluated at different time steps for HRe with a PG with primary species and a PIERSGAN, which only uses secondary species.}
	\label{dnt:fig:chievo}
\end{figure}

Bode~\cite{bode2022dpl} introduced the species splitting in the context of PG-LES. One advantage of their approach can be easily demonstrated by the non-premixed combustion case in this work. For that, a PG was trained in which all species are secondary species, i.\,e., no additional equations are solved on the reconstructed mesh. This GAN is used to advance the HRe towards the later time step discussed earlier, and the PDF of the scalar dissipation rate is shown in \reff{dnt:fig:chievo}. A PG employing primary and secondary species is able to accurately predict the PDF of the scalar dissipation rate at the later time step. The PG in which all species are secondary species is not able to predict the PDF of the scalar dissipation rate correctly, and a clear deviation is visible at about \SI{1E-3}{\per\second}. Small normally distributed errors are the reason for this deviation and accumulate over time.

\subsubsection{Effect of discriminator}
\begin{figure}[!tb]
\picsize
	\centering
    \picbox{\input{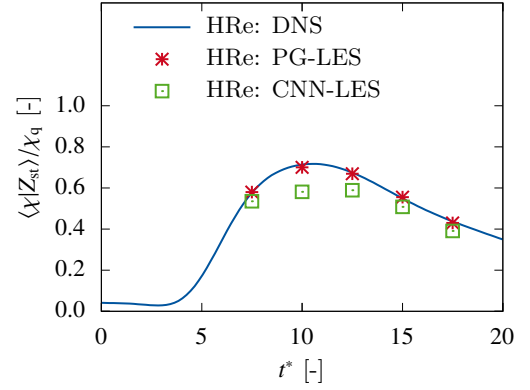}}
	\caption{Evolution of the normalized averaged scalar dissipation rate at locations with stoichiometric mixture fraction $\langle \chi\vert Z_\mathrm{st} \rangle/\chi_\mathrm{q}$ over normalized time $t^{*}$ for HRe from DNS, PG-LES, and CNN-LES.}
	\label{dnt:fig:cnn}
\end{figure}

GANs distinguish themselves from other ML/DL approaches by combining two networks during training. This has the disadvantage that it becomes more difficult to train the GAN, as the discriminator can be a source for diverging behavior. However, GANs are known to have a very good prediction accuracy. Furthermore, Bode et al.~\cite{bode2021dad} discussed how the discriminator can be used to drive the training process in situations lacking highly accurate data. Precise and fair comparisons with and without discriminator are always difficult, as the training process has an effect on the prediction quality, and it is often only optimized for one type of network. Also the comparison in \reff{dnt:fig:cnn} has this shortcoming, as the training of PG-LES, i.\,e., training of the full GAN, is highly optimized, while training of CNN-LES, i.\,e., using the generator of PG only, employed the same training strategy, which might be not perfect. As a result, the averaged scalar dissipation rate at stoichiometric mixture is slightly underpredicted and the peak is clearly shifted to a later time.  The reason is that the CNN-LES is not able to predict the level of turbulence as accurate as the PG-LES, i.\,e., it underpredicts the turbulent intensity, leading to a smaller averaged mixing rate.

\section{Application}
This section demonstrates the introduced method on two application cases. First, a PG which was trained with combined data from LRe and HRe is applied to IRe to prove the important intra-case capability of the method. Second, the amount of required training data is discussed, which is important to use PGs to accelerate simulation workflows.

\subsection{Reynolds number variation}
\label{dnt:ssec:rey}
\begin{figure}[!htb]
    \picsize
    \centering
        \begin{subfigure}[b]{22mm}
            \centering
            \includegraphics[width=\textwidth]{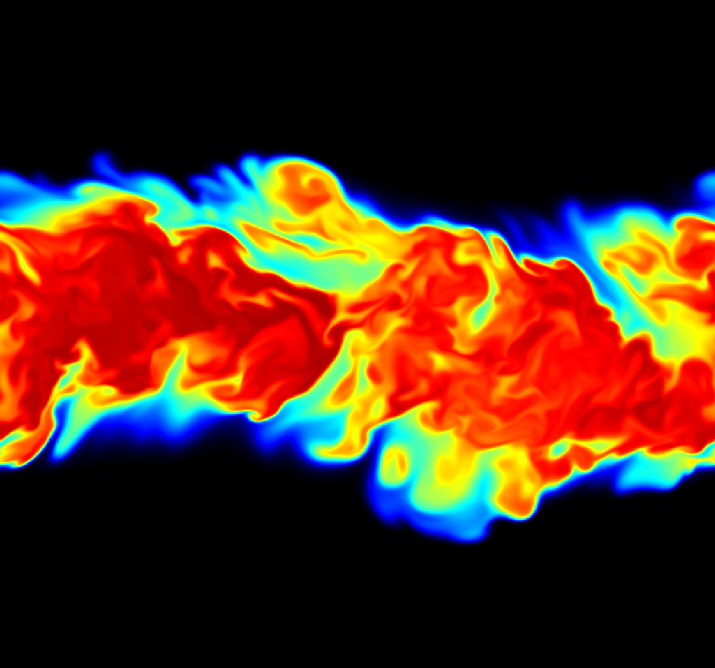}
            \vspace*{-10mm}
            \caption*{\colorbox{white}{$Z_\mathrm{H}$}}  
            \label{dnt:sfig:app6:1H}
        \end{subfigure}
        \hspace{0mm}
        \begin{subfigure}[b]{22mm}
            \centering
            \includegraphics[width=\textwidth]{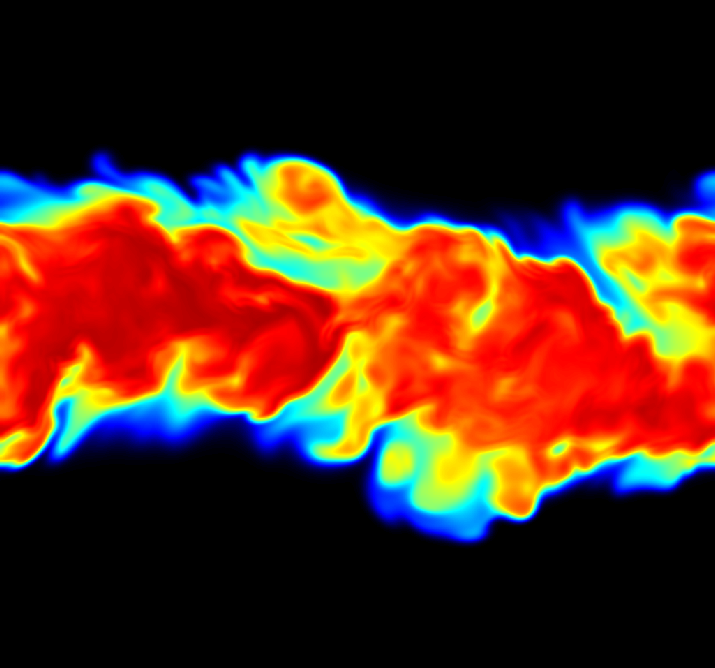}
            \vspace*{-10mm}
            \caption*{\colorbox{white}{$Z_\mathrm{F}$}}  
            \label{dnt:sfig:app6:1F}
        \end{subfigure}
        \hspace{0mm}
        \begin{subfigure}[b]{22mm}
            \centering
            \includegraphics[width=\textwidth]{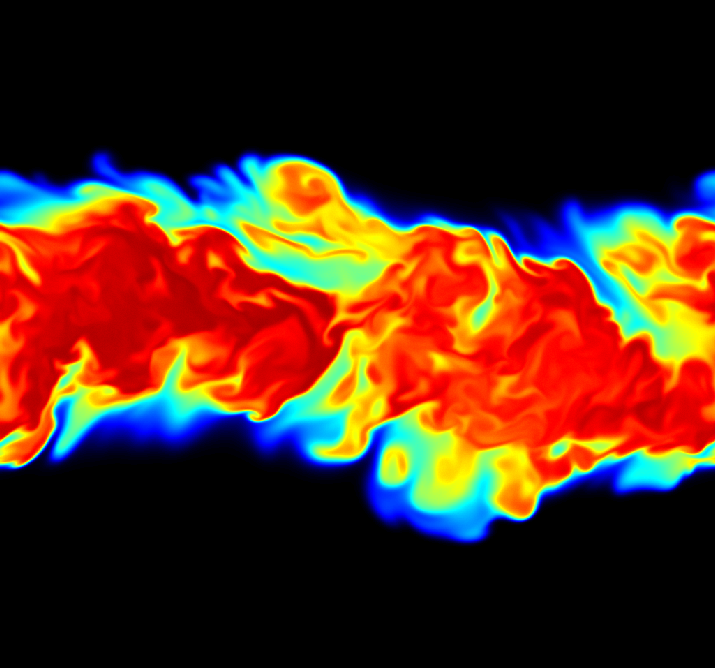}
            \vspace*{-10mm}
            \caption*{\colorbox{white}{$Z_\mathrm{R}$}}  
            \label{dnt:sfig:app6:1R}
        \end{subfigure}
        \vskip 1mm

        \begin{subfigure}[b]{22mm}
            \centering
            \includegraphics[width=\textwidth]{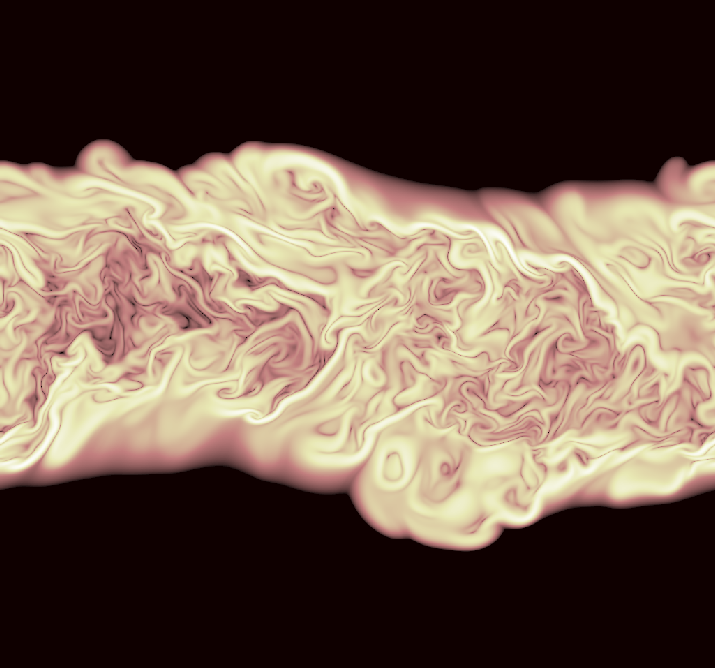}
            \vspace*{-9.5mm}
            \caption*{\colorbox{white}{$\chi_\mathrm{H}$}}  
            \label{dnt:sfig:app6:2H}
        \end{subfigure}
        \hspace{0mm}
        \begin{subfigure}[b]{22mm}
            \centering
            \includegraphics[width=\textwidth]{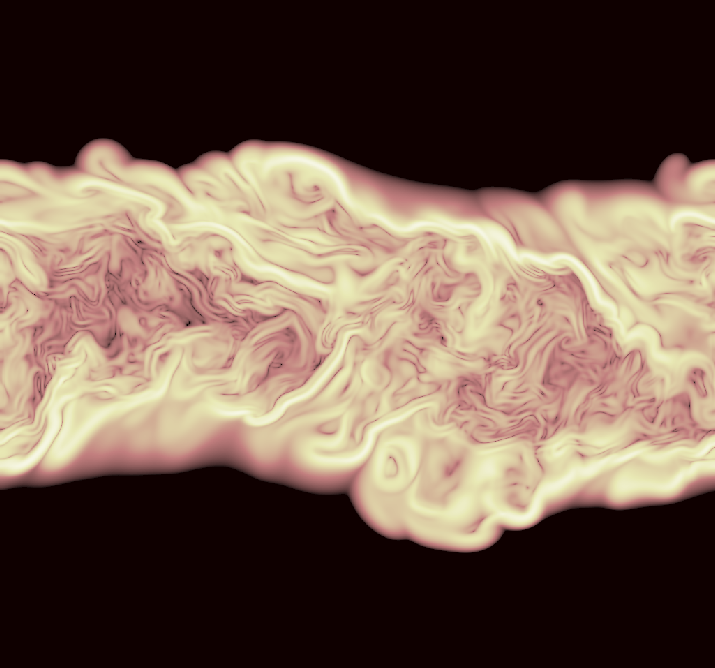}
            \vspace*{-9.5mm}
            \caption*{\colorbox{white}{$\chi_\mathrm{F}$}}  
            \label{dnt:sfig:app6:2F}
        \end{subfigure}
        \hspace{0mm}
        \begin{subfigure}[b]{22mm}
            \centering
            \includegraphics[width=\textwidth]{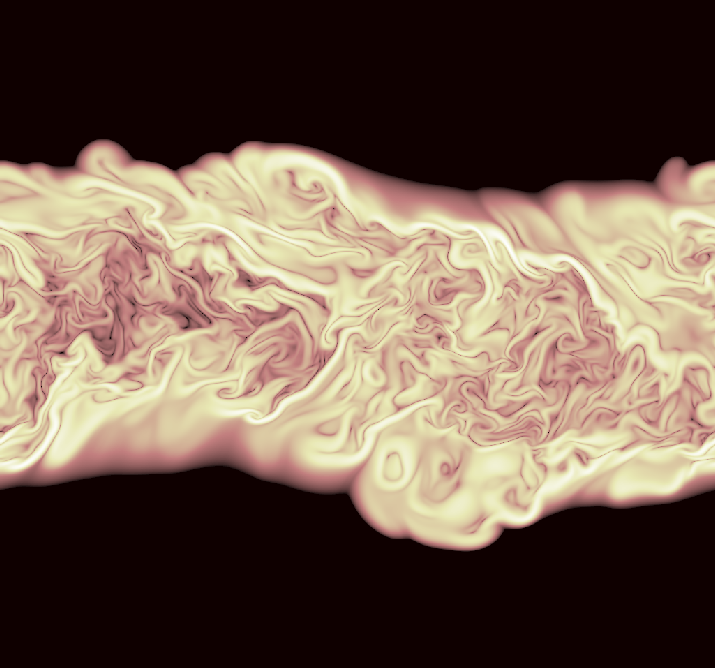}
            \vspace*{-9.5mm}
            \caption*{\colorbox{white}{$\chi_\mathrm{R}$}}  
            \label{dnt:sfig:app6:2R}
        \end{subfigure}
        \vskip 1mm

        \begin{subfigure}[b]{22mm}
            \centering
            \includegraphics[width=\textwidth]{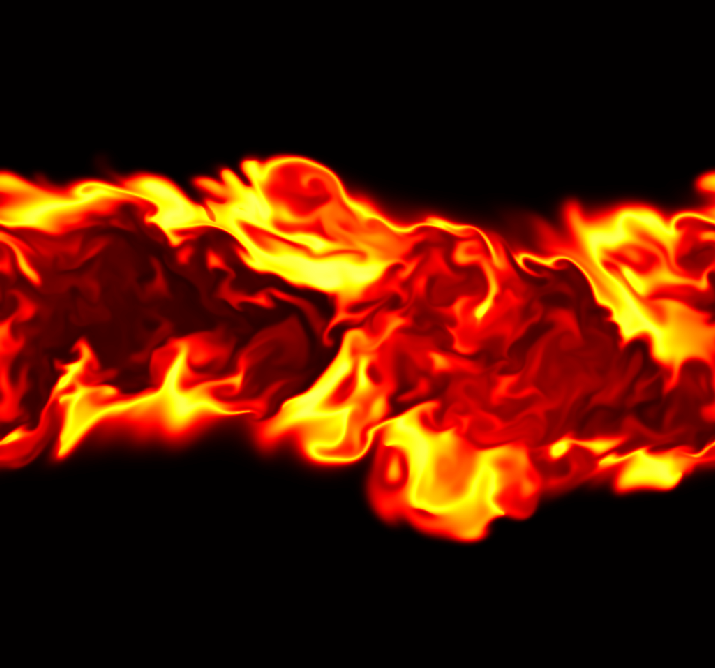}
            \vspace*{-10mm}
            \caption*{\colorbox{white}{$T_\mathrm{H}$}}  
            \label{dnt:sfig:app6:3H}
        \end{subfigure}
        \hspace{0mm}
        \begin{subfigure}[b]{22mm}
            \centering
            \includegraphics[width=\textwidth]{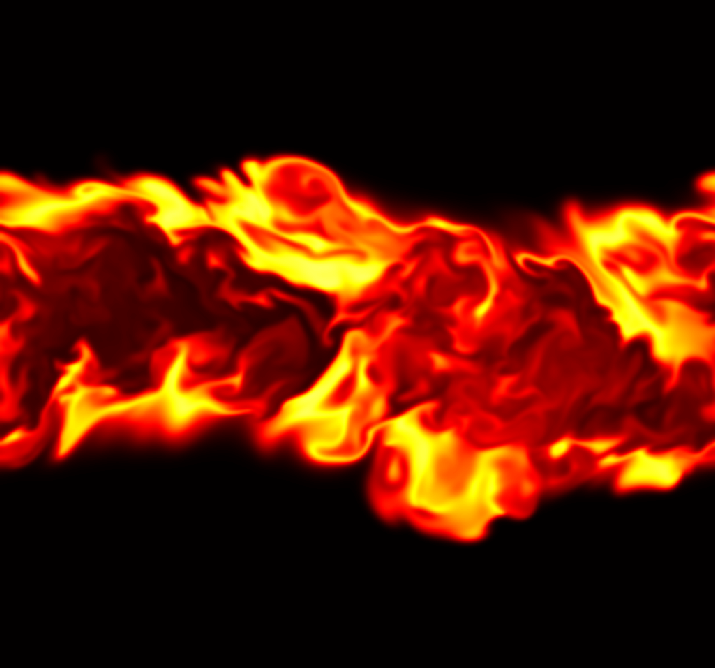}
            \vspace*{-10mm}
            \caption*{\colorbox{white}{$T_\mathrm{F}$}}  
            \label{dnt:sfig:app6:3F}
        \end{subfigure}
        \hspace{0mm}
        \begin{subfigure}[b]{22mm}
            \centering
            \includegraphics[width=\textwidth]{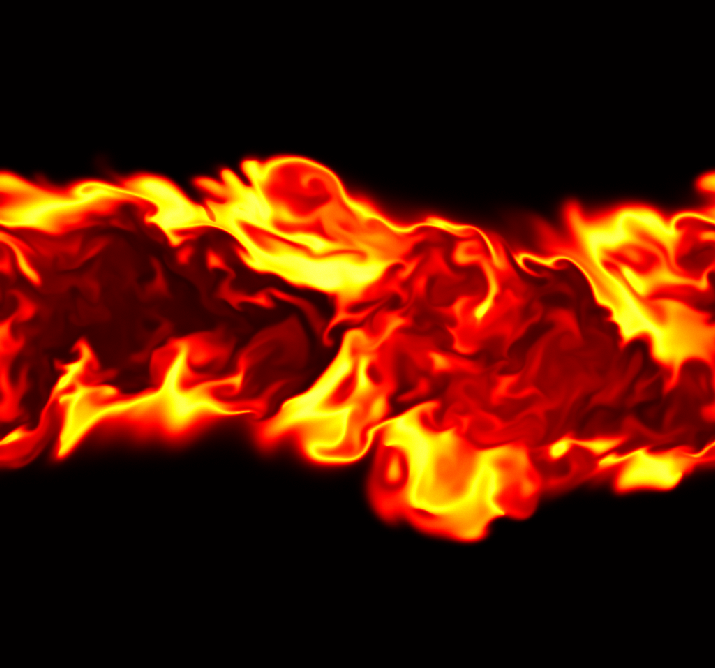}
            \vspace*{-10mm}
            \caption*{\colorbox{white}{$T_\mathrm{R}$}}  
            \label{dnt:sfig:app6:3R}
        \end{subfigure}
       \caption{Visualization of 2-D slices of fully resolved data ("H"), filtered data ("F"), and reconstructed data ("R") of the mixture fraction $Z$, the scalar dissipation rate $\chi$, and the temperature $T$ for IRe at a late time step.}
        \label{dnt:fig:app6}
    \end{figure}

To demonstrate the intra-case capabilities of the PG-LES approach, a single PG was trained with data from the LRe and HRe cases. The discriminator network was initialized from the decaying Reynolds number case by Bode et al.~\cite{bode2021dad} to better enable multiple Reynolds number. The generator of the GAN was then updated with the LRe and HRe data. The results presented in \reff{dnt:fig:app6} show that the network trained with the combined data, i.\,e., trained without the data of the actual case, gives good results in the considered a priori test.
%

\subsection{Accelerated simulation workflow}
%
The purpose of traditional LES is often to reduce the cost of simulations to make them feasible or accelerate development workflows compared to DNS. If this idea is followed for PGs, the cost for computing the training data and the training itself needs to be considered. From that point of view, it is advantageous to use PGs in scenarios which require many LES realizations. The more realizations are needed to determine a statistical effect, the better in terms of computing cost for PG, as cost for training data and training only occur once. Bode et al.~\cite{bode2022dpt} used PG to evaluate cycle-to-cycle variations (CCVs) in premixed kernels under engine conditions and demonstrated this advantage.

\begin{figure}[!tb]
\picsize
	\centering
    \picbox{\input{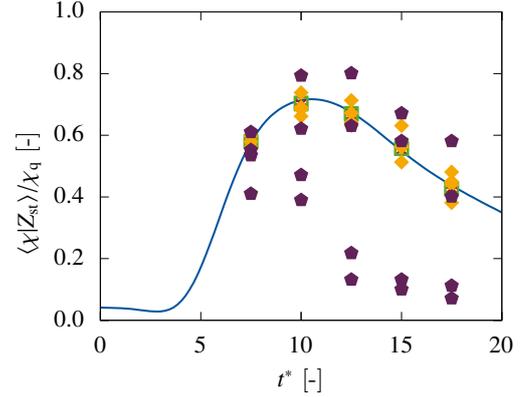}}
	\caption{Evolution of the normalized averaged scalar dissipation rate at locations with stoichiometric mixture fraction $\langle \chi\vert Z_\mathrm{st} \rangle/\chi_\mathrm{q}$ over normalized time $t^{*}$ for HRe from DNS and PG-LES trained with reduced domain sizes. Purple pentagons represent training with \SI{5}{\%},  yellow diamonds training with \SI{10}{\%}, and green boxes training with \SI{20}{\%}}
	\label{dnt:fig:tcomp}
\end{figure}

\begin{figure}[!tb]
\picsize
	\centering
    \picbox{\input{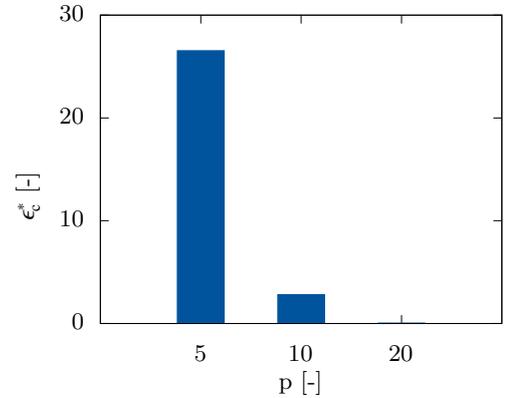}}
	\caption{Visualization of normalized error $\epsilon_\mathrm{c}^{*}$ for PG-LES trained with \SI{5}{\%}, \SI{10}{\%}, and \SI{20}{\%} domain data.}
	\label{dnt:fig:conv}
\end{figure}

The high prediction accuracy of PGs demonstrated in this work but also shown earlier \cite{bode2021dad,bode2022dlc,bode2022dpl,bode2022dpt} motivates another way to accelerate simulations, as will be discussed by means of the HRe case. The HRe case needs a very large domain with $1280\times960\times960$ cells for multiple reasons. First, the local resolution needs to fully resolve the turbulent structures and flame. Second, the boundaries in $x_2$-direction needs to be distant enough to not disturb the flow. Finally, the $x_1$- and $x_3$-directions need to be large enough to fit the largest turbulent scales but also large enough to give sufficient statistics, as statistics are typically evaluated over $x_1x_3$-planes in a planar temporal jet as considered here. I.\,e., compared to a spatially evolving jet, which increases the statistical accuracy by running longer in time, the temporally evolving jet improves its statistics by increasing the number of cells in the domain. This raises the question whether the case used for training also needs to be large for statistical reasons or whether it can be run with a reduced domain, as the statistics are finally computed on the PG-LES data, which mimic larger domains. To evaluate this question, the original setup of the HRe case was recomputed with a reduced domain size and consequently reduced number of cells in $x_2$-direction.  Reductions to \SI{5}{\%}, \SI{10}{\%}, and \SI{20}{\%} of the original case were considered.  The resulting workflow is: Compute a reduced case. Use the reduced case to train the PG network. Run PG-LES on a mesh corresponding to the DNS setup at its original size and compare the results. This is shown in \reff{dnt:fig:tcomp}. Four different realizations were trained for \SI{5}{\%} and \SI{10}{\%}, and two different realizations were trained for \SI{20}{\%}. The \SI{5}{\%}-PG-LES is not able to accurately reproduce the DNS data. The very slim DNS setup affects the turbulent field in the simulation, leading to a clear underprediction of the local turbulence intensity. Note that especially for the \SI{5}{\%} training, it is difficult to achieve PG-LES which do not blow up, however, only realizations which did not blow up were considered for \reff{dnt:fig:tcomp}. The results based on the \SI{10}{\%} training domain are already not too bad and mostly predict the trends correctly. For \SI{20}{\%}, the results converge towards the results achieved with the full training domain. This becomes also clear from \reff{dnt:fig:conv}, which shows the over all data points shown averaged error per case compared to the DNS data relatively to the error from the case trained with the full domain, defined as
\begin{equation}
\epsilon_\mathrm{c}^{*}=\frac{|\delta_p|-|\delta_{100}|}{|\delta_{100}|}
\end{equation}
with $\delta$ as deviation of each PG-LES data point from the DNS data points and $p$ as domain size in percentage, i.\,e., $100$ denotes the full DNS domain.

\section{Conclusions}
This paper extends the methodology of PG-LESs and applies PG-LES to a non-premixed combustion temporal jet case. The presented generalization with respect to underlying meshes shows high accuracy and is especially important for target cases with complex geometries. The given a priori and a posteriori results emphasize the high prediction quality of the PG-LES approach which is remarkable considering the complexity of the model finite-rate-chemistry flow. Overall, PG-LES seems to be a very promising approach to accelerate the development of future flow-driven applications, such as the next generation of turbines and engines.

As the prediction quality of PG-LES was shown to be already high, the remaining main issue is the complexity of training suitable networks. Currently, there is no guarantee that a trained network gives good results, and typically multiple networks need to be trained to find one well-working one. New approaches simplifying this procedure would be desired. Moreover, limits in terms of extrapolation capabilities of GANs should be addressed in more detail in the future.

\section*{Acknowledgements}
The author acknowledges computing time grants for the projects JHPC55 and TurbulenceSL by the JARA-HPC Vergabegremium provided on the JARA-HPC Partition part of the supercomputer JURECA at J\"ulich Supercomputing Centre, Forschungszentrum J\"ulich,  the Gauss Centre for Supercomputing e.V. (www.gauss-centre.eu) for funding this project by providing computing time on the GCS Supercomputer JUWELS at Jülich Supercomputing Centre (JSC), and funding from the European Union's Horizon 2020 research and innovation program under the Center of Excellence in Combustion (CoEC) project, grant agreement no. 952181. 



\bibliographystyle{elsarticle-num} 
\bibliography{literature}


%


\end{document}